\documentclass[fleqn,usenatbib]{mnras}

\usepackage[T1]{fontenc}

\usepackage{longtable}
\DeclareRobustCommand{\VAN}[3]{#2}
\let\VANthebibliography\thebibliography
\def\thebibliography{\DeclareRobustCommand{\VAN}[3]{##3}\VANthebibliography}

\newcommand{\se}{{\sc SExtractor\ }}

\newcommand{\dos}{{\sc SExtractor+PSFEx\ }}

\usepackage{graphicx}	% Including figure files
\usepackage{amsmath}	% Advanced maths commands
\usepackage{amssymb}	% Extra maths symbols
\usepackage{tikz}
\usetikzlibrary{positioning,calc}
\usepackage{multirow}
\usepackage{adjustbox}
\usepackage[graphicx]{realboxes}

\title[The VVV near-IR galaxy catalogue in a Northern part of the Galactic disc]{The VVV near-IR galaxy catalogue in a Northern part of the Galactic disc}

\author[ Daza-Perilla et al. ]
{
\parbox[t]{\textwidth}{ I. V. Daza-Perilla  $^{1,2}$, M. A. Sgr\'o$^{1,3}$, L. D. Baravalle$^{1,3}$, M. V. Alonso$^{1,3}$,  C. Villalon$^{1}$, M. Lares$^{1,3}$, M. Soto$^{6}$, J. L. Nilo Castell\'on$^{4,5}$, C. Valotto$^{1,3}$, P. Marchant Cort\'es$^{4,5}$, D. Minniti$^{7,8,9}$ \& M. Hempel$^{7,10}$ }
\vspace*{6pt} \
\\
$^{1}$ Instituto de Astronom\'ia Te\'orica y Experimental, CONICET-UNC, C\'ordoba, X5000BGR, Argentina.\\
$^{2}$ Facultad de Matem\'atica, Astronom\'ia, F\'isica y Computaci\'on, Universidad Nacional de C\'ordoba (UNC), C\'ordoba,\\
CP:X5000HUA, Argentina.\\
$^{3}$ Observatorio Astron\'omico de C\'ordoba, Universidad Nacional de C\'ordoba, Laprida 854, C\'ordoba, X5000BGR, Argentina.\\
$^{4}$ Instituto Multidisciplinario en Investigación y Postgrado (IMIP), Universidad de La Serena. Av. Raúl Bitrán Nachary No 1305, \\
La Serena, 1720236, Chile.\\
$^{5}$ Departamento de Astronom\'ia, Universidad de La Serena. Av. Juan Cisternas 1200, La Serena, 1720236, Chile.\\
$^{6}$ Instituto de Astronom\'ia y Ciencias Planetarias, Universidad de Atacama, Copayapu 485, Copiap\'o, 1532297, Chile.\\
$^{7}$ Instituto de Astrof\'isica, Facultad de Ciencias Exactas, Universidad Andr\'es Bello, Av. Fernandez Concha 700,  \\ 
Las Condes, Santiago, 7550000, Chile.\\
$^{8}$ Vatican Observatory, V00120 Vatican City State, Italy.\\
$^{9}$ Departamento de Fisica, Universidade Federal de Santa Catarina, Florianópolis, Santa Catarina, 88040 970, Brasil.\\
$^{10}$ Max Planck Institute for Astronomy, Königstuhl 17, 69117 Heidelberg, Germany.
}

\date{Accepted XXX. Received YYY; in original form ZZZ}

\pubyear{2023}

\begin{document}
\label{firstpage}
\pagerange{\pageref{firstpage}--\pageref{lastpage}}
\maketitle

% Abstract of the paper
\begin{abstract}
The automated identification of extragalactic objects in  large surveys  provides reliable and reproducible samples of galaxies in less time than procedures involving human interaction.  However, regions near the Galactic disc are more challenging due to the dust extinction.   We present the methodology for the automatic classification of galaxies and non-galaxies at low Galactic latitude regions using both images and, photometric and morphological near-IR data from the VVVX survey. Using the VVV-NIRGC, we analyse by statistical methods the most relevant features for galaxy identification.  This catalogue was used to train a CNN with image data and an XGBoost model with both photometric and morphological data and then to generate a dataset of extragalactic candidates. This allows us to derive probability catalogues used to analyse the completeness and purity as a function of the configuration parameters and to explore the best combinations of the models. 
As a test case, we apply this methodology to the Northern disc region of the VVVX survey, obtaining 172,396 extragalatic candidates with probabilities of being galaxies.  We analyse the performance of our methodology in the VVV disc, reaching an F1-score of 0.67, a 65 per cent purity and a 69 per cent completeness. We present the VVV-NIR Galaxy Catalogue: Northern part of the Galactic disc comprising 1,003 new galaxies, with probabilities greater than 0.6 for either model, with visual inspection and  with only 2 previously identified galaxies. In the future, we intend to apply this methodology to other areas of the VVVX survey.
\end{abstract}

% Select between one and six entries from the list of approved keywords.
% Don't make up new ones.

%%%%%%%%%%%%%%%%%%%%%%%%%%%%%%%%%%%%%%%%%%%%%%%%%%

%%%%%%%%%%%%%%%%% BODY OF PAPER %%%%%%%%%%%%%%%%%%

%%%%%%%%%%%%%%%%%%%%%%%%%%%%%%%%%%%%%%%%%%%%%%%%%

%%%%%%%%%%%%%%%%%%%%%%%%%%%%%%%%%%%%%%%%%%%%%%%%%%%%%%%%
%%%%%%%%%%%%%%%%%%%%%%%%%%%%%%%%%%%%%%%%%%%%%%%%%%%%%%%%
%%%%%%%%%%%%%%%%%%%%%%%%%%%%%%%%%%%%%%%%%%%%%%%%%%%%%%%%

\section{Introduction}\label{sec:Introduction}

The use of statistical tools capable of automatically generating models, deriving catalogues and determining the statistical description of the physical quantities of stellar objects has been favoured by the large amounts of available data. One of these tools is the availability of mathematical models that can automatically learn from data and that, albeit more complex than the classical methods, provide the opportunity to solve various tasks using the large amount of data from surveys \citep[see, e.g., ][]{2019arXiv190407248B}. 

The types of problems that can be solved with machine learning approaches include, but are not limited to, solutions via unsupervised models such as the detection of anomalies in the Sloan Digital Sky Survey (SDSS) quasar spectra \citep{2010AJ....140..390B}, the dimensionality reduction of infrared spectra for the determination of some physical characteristics of stars \citep{2018MNRAS.476.2117R} and the visual inspection of data through the embedding of features into another variables space \citep{Reis21}. In terms of supervised models for regression or classification, for instance, the morphological classification of galaxies \citep{Spindler+21}, Young Star Object finders \citep[][]{Marton+2019}, automated classification of eclipsing binary systems in the VVV Survey \citep{2023Daza}, Drifting  Features: detection and evaluation in the context of automatic RRLs identification in VVV \citep[][]{2021A&A...652A.151C}, variable star classification across the Galactic bulge and disc with the VISTA Variables in the Vía Láctea survey \citep[][]{2022MNRAS.509.2566M} with light curves of stars, classification of galaxies and QSOs \citep{Logan+20}, estimation of photometric redshifts \citep{2020MNRAS.497.4565E}, among many others.  

The detection and identification of extragalactic sources at low Galactic latitudes are crucial for understanding the distribution of galaxies across the sky. However, these detections are harder than elsewhere because they are strongly influenced by the presence of gas, dust and high stellar concentration towards the Galactic disc \citep{2019MNRAS.482.5167S}.
In the Zone of Avoidance (ZoA, \citealt{Kraan2018}), where optical wavelengths are influenced by high Galactic extinction, the use of infrared passbands made possible the exploration of this region via the Two Micron All Sky Survey (2MASS; \citealt{Skrutskie2006}). With these data, \cite{2000AJ....119.2498J} identified galaxy candidates in these crowded regions behind the Milky Way.   

The VISTA Variables in the Vía Láctea (VVV, \citealt{Minniti2010}) is a deeper near-infrared (NIR) survey of the Hi, Galactic bulge and Southern part of the disc. Its main scientific objective was the study of Galactic variable stars but to also reveal galaxies behind our Galaxy. The extension of this survey that triples the sky coverage in the ZoA is the VISTA Variables in the Vía Láctea eXtended Survey (VVVX, \citealt{Minniti2018}). It includes a Northern part of the disc and increases notably the Southern part. Using data from these surveys, several works have been carried out for galaxy detections \citep[see, ][]{Amores2012, Coldwell2014, Baravalle2018, Baravalle2019, Galdeano2021, Galdeano2022}. With the disc data of the VVV survey, \cite{Baravalle2021} presented the VVV NIR Galaxy Catalogue (VVV NIRGC), this catalogue of galaxies in these regions with visual inspection. 

A photometric and morphological procedure capable of performing a separation between point and extended sources was carried out. Due to the increased amount of available data in the VVVX survey, it would be too time consuming or nearly impossible to inspect all the galaxy candidates in these extended regions of the disc. This puts a severe constraint in our original procedure, but fortunately in a more efficient way machine learning techniques allow us to discr
iminate and detect galaxies. We designed a procedure capable of performing an automatic identification of galaxies and non-galaxies in the VVV and VVVX survey regions. We use statistical methodologies including unsupervised and supervised machine learning techniques implemented on images and photometric information independently.

This work also has interesting future projections with the advent of the NASA Nancy Grace Roman Space Telescope \citep[a.k.a. WFIRST,][]{2012arXiv1208.4012G, 2015arXiv150303757S}.
This is a near-IR survey telescope to be launched in $\sim$ 2026 whose wavelength coverage has been extended to contain the $K_s$-passband filter \citep[][]{2018arXiv180600554S}.
A near-IR survey of the Galactic plane with the Roman Space Telescope would reach $\sim$ 4 mag deeper than the VVV survey images (R. Paladini et al. 2022, private communication).
In addition, the ``R2D2 synergy'' of the Roman Space Telescope with the Vera C. Rubin Observatory Legacy Survey of Space and Time \citep[][] {2009arXiv0912.0201L,  2019ApJ...873..111I} will have similar sensitivities, as well as complementary optical wavelengths, spatial resolutions, and time coverage \citep[][]{2022arXiv220212311G}.
The combination of both facilities would not only make a deep map of the distribution of stars and dust in the Galactic plane, but also reveal what is beyond the Milky Way.
The present analysis may then be extended to study much fainter galaxies and underlying large scale structure in the regions of the ZoA analysed here.

The paper is organised as follows. Section~\ref{sec:Data} explains the data used in this work. Section~\ref{sec:feature_selection} describes the methods for selecting the most important features for galaxy identification.  Section~\ref{sec:Identifying_galaxies} contains the methodology for the identification through supervised methods trained and applied in the disc of the VVV survey. In Section~\ref{sec:catalogue}, we present the identification and a new catalogue of galaxies in the Northern disc regions. Finally, Section~\ref{sec:conclusions} presents the summary and conclusions of this work.

\section{Data}\label{sec:Data}
 
The VVV survey comprises the Galactic bulge and a large part of the Southern Galactic disc (\citealt{Minniti2010}). 
The VVVX (\citealt{Minniti2018}) is an extended survey that follows the same observational strategy used for the VVV survey in three passbands: $J$ (1.25 $\mu$m), $H$ (1.64 $\mu$m), and $K_{s}$ (2.14 $\mu$m) as well as variability information for the $K_{s}$ passband. Both surveys were divided in tiles of 1$^{\circ}$  $\times$ 1.5$^{\circ}$ produced by six single pointing observations. Table~\ref{tab:table1} shows the basic area description of the VVV and VVVX surveys, listing in column (1) the observed regions, in columns (2) and (3), the Galactic coordinates and in column (4), the number of tiles in each region.  Through this work, we refer to the disc+20 and disc observed regions of Table~\ref{tab:table1} as the Northern and VVV disc, respectively.

\begin{table}
\center
\caption {The VVV and VVVX surveys: areal coverage and number of tiles.  Here Northern part of the Galactic disc is called as disc+20. }

\begin{tabular}{lccr}
\hline
Observed &  Galactic &  Galactic & Number \\
region   & longitude &  latitude & of tiles \\
\hline
VVV survey &     &   & \\
\hline
bulge      &  350$^{\circ}$  $<$ $l$ $<$ 10 $^{\circ}$   &  -10$^{\circ}$ $<$ $b$ $<$ +5$^{\circ}$  & 196 \\
disc      &  295$^{\circ}$ $<$ $l$ $<$ 350$^{\circ}$    &  -2.25$^{\circ}$ $<$ $b$ $<$  +2.25$^{\circ}$ & 152\\
\hline
VVVX survey &     &   & \\
\hline
bulge-low  &  350$^{\circ}$ $<l<$ 10$^{\circ}$    &  -15$^{\circ}$   $<b<$ -10$^{\circ}$   & 56 \\
bulge-high &  350$^{\circ}$ $<l<$ 10$^{\circ}$    &  +5$^{\circ}$    $<b<$ +10$^{\circ}$   & 56 \\
disc+20    &  10$^{\circ}$ $<l<$ 20$^{\circ}$     &  -4.5$^{\circ}$  $<b<$ +4.5$^{\circ}$  & 56 \\
disc-low   &  230$^{\circ}$ $<l<$ 350$^{\circ}$   &  -4.5$^{\circ}$  $<b<$ -2.25$^{\circ}$ & 166 \\
disc-high  &  230$^{\circ}$ $<l<$ 350$^{\circ}$   &  +2.25$^{\circ}$ $<b<$ +4.5$^{\circ}$  & 166 \\
disc+230   &  230$^{\circ}$ $<l<$ 295$^{\circ}$   &  -2.25$^{\circ}$ $<b<$ +2.25$^{\circ}$ & 180 \\
\hline
\label{tab:table1}
\end{tabular}
\end{table}

In \cite{Baravalle2021}, they used the images generated by the VVV survey in the 152 tiles of the Southern part of Galactic disc, which are available via the Archive Science Portal or programmatically\footnote{\href{https://archive.eso.org/scienceportal/home?data_release_date=*:2021-10-21&data_collection=VVVX&publ_date=2021-10}{archive.eso.org}}. These images were selected with observed status 'Completed', the same exposure time and seeing $<$~0.9~arcsec in the $K_{s}$ passband. They applied a pipeline, which contains the morphological and photometric procedure described in \cite{Baravalle2018} to identify and classify sources. 
Implementing \dos on the $J$, $H$ and $K_{s}$ images they detected 177,838,607 objects. Using the pipeline, they discarded point sources and obtained 2,070,768 extended objects. After the colour selection, they obtained a sample of 80,522 possible extragalactic sources. In this work, we performed a cross-match with Gaia-DR3 \citep[][]{2021A&A...650C...3G} using a separation of 2 arcsec between sources. We found 258 common sources that were removed from our analysis for being considered stellar objects.  We discarded the objects situated at the edges of the tiles and eliminated any duplicate sources.  %The steps of this improved pipeline are summarised in the flowchart of Figure~\ref{fig:Flowchart1} presented in Appendix~\ref{app:pipeline}. 

\begin{figure*}
\centering
\includegraphics[width=1\textwidth]{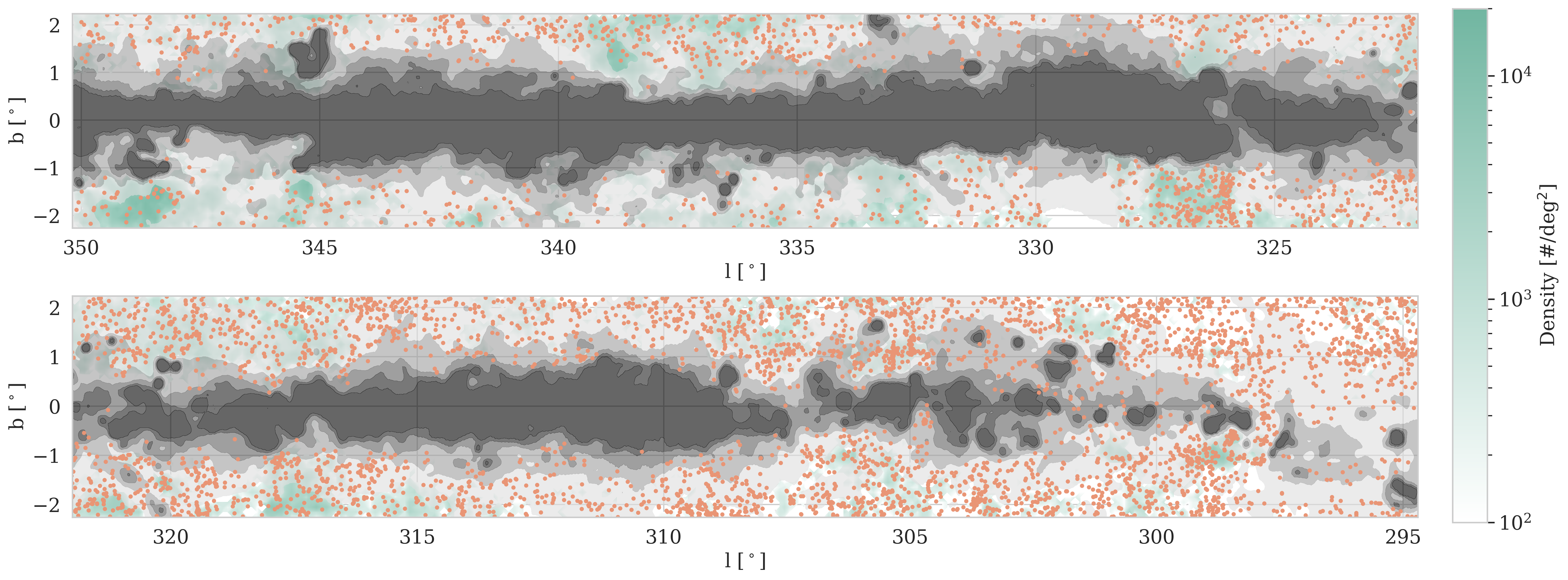} 
\caption{Southern Galactic disc with Galactic longitudes between 350$^\circ$ and 295$^\circ$. Gxs are represented by orange points and non-Gxs as a density green-scale map. The total optical A$_{V}$ interstellar extinction from the maps of \citet{Schlafly2011} is superimposed in a grey gradient with levels of 10, 15, 20, and 25 mag.} 
\label{fig:CrossMatch}
\end{figure*}

The  resulting sample of possible extragalactic sources on the VVV disc consists of 80,038 objects. These objects were visually inspected in the process of generating the VVV NIRGC catalogue \citep{Baravalle2021}. Therefore, taking advantage of this visual inspection or, in other words, this labelling of objects, we obtained two samples, 5,509 objects that are galaxies (hereafter, the Gx sample) and 74,238 non-galaxies (hereafter, the non-Gx sample), which we used for the construction of the identifier|classifier.  Figure~\ref{fig:CrossMatch} shows the two samples: Gxs as points and non-Gxs as the number of points per square degree of the Southern part of the disc of our Galaxy. 
Also superimposed is the total optical A$_{V}$ interstellar extinction from the maps of \citet{Schlafly2011} in a grey gradient with levels of 10, 15, 20, and 25 mag.

The improved pipeline was also applied on the 56 tiles of the Northern Galactic disc of the VVVX. We chose this region due to its similarities in interstellar extinction to the Southern disc of the VVV survey. We detected 66,983,004 objects, of which 172,396 are possible extragalactic sources. Our main goal was to classify these sources as Gx and non-Gx with a certain probability given by supervised machine learning models and to select those with higher probability in order to generate a new catalogue of galaxies in the Northern Galactic disc.

\subsection{Samples}

The performance of machine learning models depends strongly on the amount and type of information provided. For the detection of galaxies and the classification of possible extragalactic sources into Gxs and non-Gxs with machine learning techniques, we used two independent approaches, images and photometric information.
The reasons for the split are: i) the possible difficulty in acquiring both types of information for a particular source; ii) double confirmation of the classification of supervised machine learning models trained with image information and physical object information; iii) the computational cost in the classification and; iv) the estimation of the quality of the results when using image data, photometry, or both. For this, we constructed samples from the set of possible extragalactic sources as described below.

\subsubsection{Image-based Samples}

The images of the sources are a collection of pixel intensities contained in a matrix of a given size. The samples using this image information are hereafter referred to as IS.  The median intensities of the images in the $J$, $H$ and $K_s$ passbands of the whole sample are 1827.75, 9333.25 and 12680.5 ADUs, respectively. We tested different matrix sizes using the tiles of the disc region of the VVV survey in the $J$, $H$ and $K_{s}$ passbands. We built sets of images with different spatial sizes, one of them similar to the visual classification of \cite{Baravalle2021} and the others with a smaller size trying to cover a smaller fraction of sky background and stellar contamination but keeping a good part of the centre of the object to be classified: $\sim$ 15 $\times$ 15 arcsec equivalent to 44 $\times$ 44 pixels; $\sim$ 4.407 $\times$ 4.407 arcsec equivalent to 13 $\times$ 13 pixels; 3.729 $\times$ 3.729 arcsec equivalent to 11 $\times$ 11 pixels and $\sim$ 3.051 $\times$ 3.051 arcsec equivalent to 9 $\times$ 9 pixels.

\subsubsection{Photometry-based Samples}\label{subsec:PS}

In what follows, the sample of objects with photometric information is referred to as PS\footnote{Photometric Sample.}. For the Gxs, we extracted the information from the VVV NIRGC catalogue \citep{Baravalle2021} and for the non-Gxs, we used the output of the pipeline obtained with \dos.

\begin{figure*}
    \centering
    \includegraphics[width=1.0\linewidth]{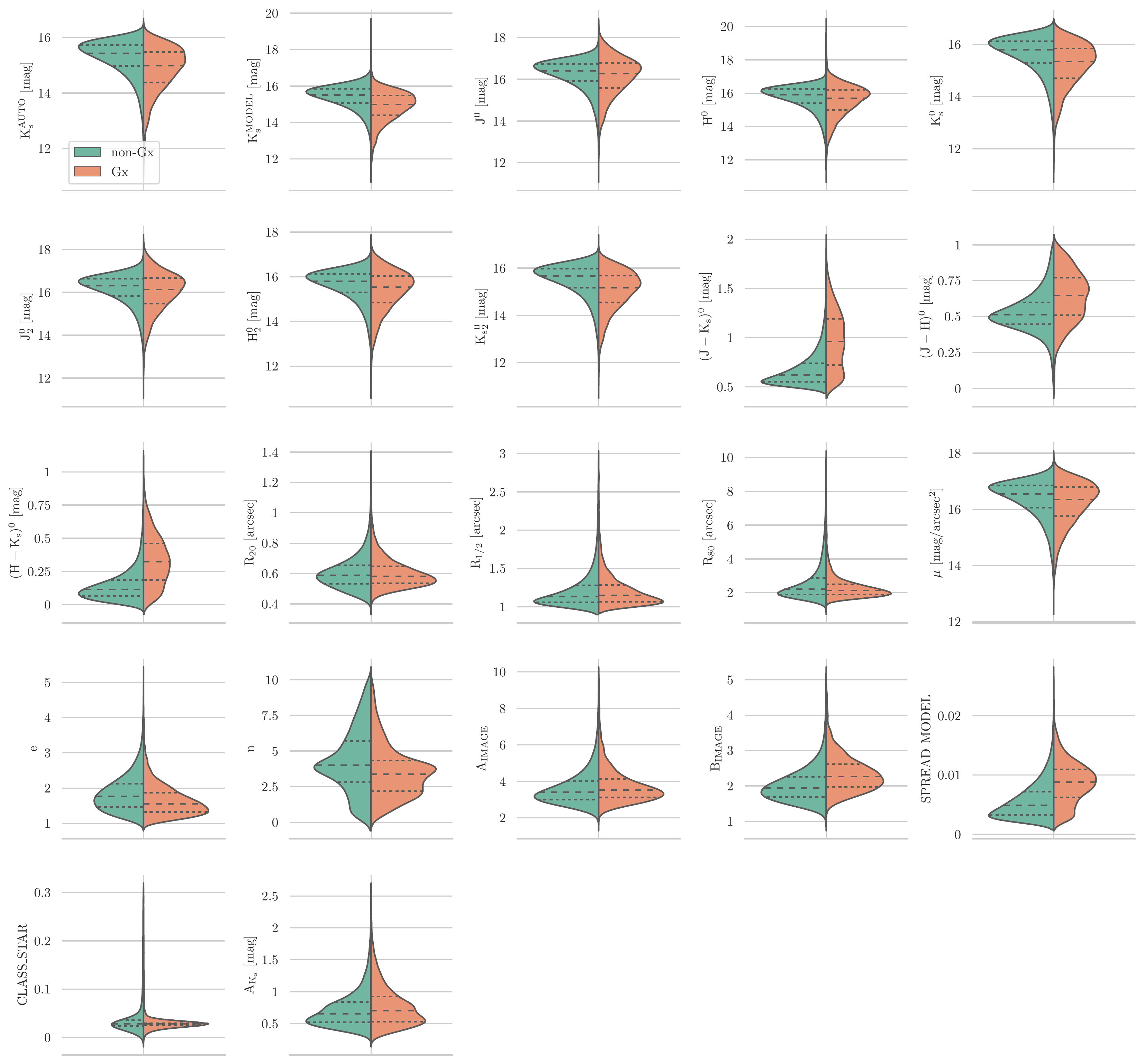}
    \caption{Kernel density estimation of the features in the PS sample. The green distributions correspond to the objects visually labelled as non-Gxs, while the orange ones, to the Gxs. The dashed lines mark the sample medians and the dotted lines are the confidence intervals.}
    \label{fig:all_varia_violin}
\end{figure*}

For each object in the PS sample, the data set contains photometric (namely, total extinction-corrected $J^{0}$, $H^{0}$, and $K_{s}^{0}$ magnitudes; the extinction-corrected  $J_{2}^{0}$, $H_{2}^{0}$ and $K_{s}{}_{2}^{0}$ aperture magnitudes within a fixed aperture of 2 arcsec diameter; aperture ($J$ -$K_{s}$)$_{2}^{0}$, ($J$ - $H$)$_{2}^{0}$ and ($H$ - $K_{s}$)$_{2}^{0}$ corrected colours; surface brightness, $\mu$; AUTO and MODEL magnitudes in the $K_{s}$ passband) and morphological information (half-light radius, $R_{1/2}$; $20\%$-light radius, $R_{20}$; $80\%$-light radius, $R_{80}$; semi-major and semi-minor axes of the isophotal, $A_{\textrm{IMAGE}}$ and $B_{\textrm{IMAGE}}$ respectively; elongation, $e$ and spheroid Sersic index, $n$). 
We also included those variables computed by \se and used in the pipeline when selecting galaxy candidates (CLASS\_STAR and SPREAD\_MODEL) and A$_{K_s}$ interstellar extinction in the $K_{s}$ passband where the object is located. Figure~\ref{fig:all_varia_violin} shows the kernel density estimation of the underlying distribution corresponding to objects visually labelled as Gxs and non-Gxs.

\subsection{Training and test split}

In order to perform a fair comparison between the models trained with the IS and PS samples, we defined two subsets of data selecting the same objects. We obtained a training and a test set containing 70\% and 30\% of the total number of objects, respectively. Since our data have an imbalance of $\approx 1:13$ in the target variable, these subsets are stratified by this variable to preserve this imbalance.

The training set is used for the analysis of relevant information to galaxy identification and the choice of the classifier hyper-parameters. Meanwhile, the test set is used to determine the confidence in the model performing the task, in our case the generation of galaxy catalogues.  The imbalance of the classes corresponding to the Gx and non-Gx samples in the training and test sets are shown in Table~\ref{tab:table_2}.

\begin{table}
    \caption{Class balance for the possible extragalactic sources (IS | PS) of the total set in the first row and of the training and test sets in the second and third rows, respectively.}
    \centering
    \begin{tabular}{lccc}
    \hline
    Samples &  Gx & Non-Gx & Total\\
    \hline
    Possible extragalactic sources & 5,509 & 74,283 & 79,792\\
    Training                       & 3,856 & 51,998 & 55,854\\
    Test                           & 1,653 & 22,285 & 23,938\\
    \hline
    \end{tabular}
    \label{tab:table_2}
\end{table}

\section{Feature selection}\label{sec:feature_selection}

One of the most important steps in the classifier determination is the generation and selection of the features with the highest entropy. To do this, we implemented different statistical methods, in particular unsupervised learning and univariate analysis. In the unsupervised learning, there are algorithms that work with unlabelled data, such as \textit{k}-means, Voronoi tesselation, Gaussian Mixture Model, and density-based spatial clustering of applications with noise (DBSCAN, \citealt{DBSCAN, Masters+17, Masters+19, Hemmati+19, Logan+20}).

For this study, we use the \textit{k}-means algorithm as an unsupervised  method. The \textit{k}-means algorithm consists of dividing a set of {\it N} samples {\it X} into {\it K} disjoint clusters {\it C}, each described by the mean $\mu_j$ of the samples in the cluster. The means are commonly called the cluster “centroids”; each observation {\it X} belongs to the cluster with the nearest mean (equation~\ref{equ:1}).

\begin{equation}
    \hspace{2.1 cm}\sum_{i=0}^{n} min {||x_i - \mu_j||}^2\hspace{0.7 cm}; \hspace{0.7 cm}\mu_j \in C 
    \label{equ:1}
\end{equation}

Regarding univariate analysis methods, we use  principal component analysis \citep[PCA,][]{Hotelling1933AnalysisOA} and mutual information \citep[MI,][]{6773024}. The first is employed in exploratory data analysis and for predictive models. It is also commonly used for dimensionality reduction, projecting each data point onto the first principal components, such that the variance of the data is preserved as much as possible. The first principal component can equivalently be defined as a direction that maximises the variance of the projected data. The second principal component can be taken as a direction orthogonal to the first principal components that maximises the variance of the projected data. On the other hand, the MI method is a non-negative value, which  measures the nonlinear dependence between two random variables by quantifying the amount of information that can be obtained about one of them by observing the other. It is equal to zero if and only if two random variables are independent, and higher values mean higher dependency.

\subsection{Image-based Samples}

The size of the images and the number of passbands are used to define the number of features that an image might have. In the case of an image of 11 $\times$ 11 pixels in the three passbands $J$, $H$ and $K_s$, each pixel can be considered as a feature. Therefore the feature space where each image of 11 $\times$ 11 $\times$ 3 lies, has a dimension of 363. In this space, where each feature is a dimension, we analyse the most important features for the classification of Gxs and non-Gxs through statistical methods. If the features are informative for the distinction between Gxs and non-Gxs, the image data should show a structure in the feature space that is associated with each class: Gxs and non-Gxs. As the distribution of the representative points of each image in this space of dimensions is higher than three, it cannot be visualised. To study the existence of some intrinsic structure based on the intensity information contained in the images, we used the PCA and \textit{k}-means methods with two components and k equal to two, respectively.

\subsubsection{Generation}

In the process of feature selection, we studied the importance of feature generation through passband differences, pixel scaling by varying the intensity distribution, and smoothing and edge detection filters. The study was tested on images of different spatial sizes: 44 $\times$ 44, 13 $\times$ 13, 11 $\times$ 11  and 9 $\times$ 9 pixels.

\textbf{Band differences:}
Passbands in relation to spectra, allow certain wavelength ranges to pass, so that an object can be measured from the point of view of several separate passbands and compared. These passbands are useful when calculating the colour of an object, which is defined using two magnitudes of an object spectrum. In general, an object colour is used to discriminate between point and extended sources or morphological types of galaxies \citep{2015AAS...22533651S}, therefore, given that colours are defined from fluxes and these from pixel intensities, we study whether the subtraction between passband intensities is relevant for the separation between Gxs and non-Gxs.

\textbf{Scaling:}
Generally, astronomical image processing does not include scaling. However, in many cases during the visual classification process of Gx and non-Gx, an RGB image is generated combining the information of the $J$, $H$, and $K_s$ passbands, where a colour is assigned to each passband and their intensities are scaled from 0 to 255. With this point in mind and knowing that many supervised machine learning models require the data to be scaled, we study what kind of scaling is suitable for the separation between Gxs and non-Gxs. 

The scaling method implemented is min-max. This method preserves the shape of the original distribution and does not significantly change the information contained in the original data. However, it does not reduce the importance of outliers. The min-max method consists of taking each value of a feature, subtracting the minimum value of a range of intensities associated with the feature and dividing by the difference between the maximum and minimum of this range. In particular, we study how the variation of the range of intensities associated to the features impacts on the identification of the Gxs. For this we compare four ways of implementing the min-max method on each image, i.e. we use four different intensity ranges:

\begin{enumerate}
     \item Scaled by image (scl\_image): each image is scaled using the intensity range of all its pixels including all passbands.
     
    \vspace{0.5 cm}
    \item Scaled by passband (scl\_band): each passband for each image is scaled independently using the intensity ranges for their own $J$, $H$ and $K_s$ pixels.
    
    \vspace{0.5 cm}
    \item Scaled by passband by set (scl\_band\_set): each passband for each image is scaled using the intensity range of all $J$, $H$, $K_s$ pixels in the whole sample.
    
    \vspace{0.5 cm}
    \item Scaled by passband by tile (scl\_band\_tile): Each image is scaled using three ranges of intensities corresponding to the pixel values of the $H$, $J$ and $K_s$ passbands, respectively, but in this case considering all the images belonging to a tile.
\end{enumerate}

\begin{figure*}
    \centering
    \includegraphics[width=1\linewidth]{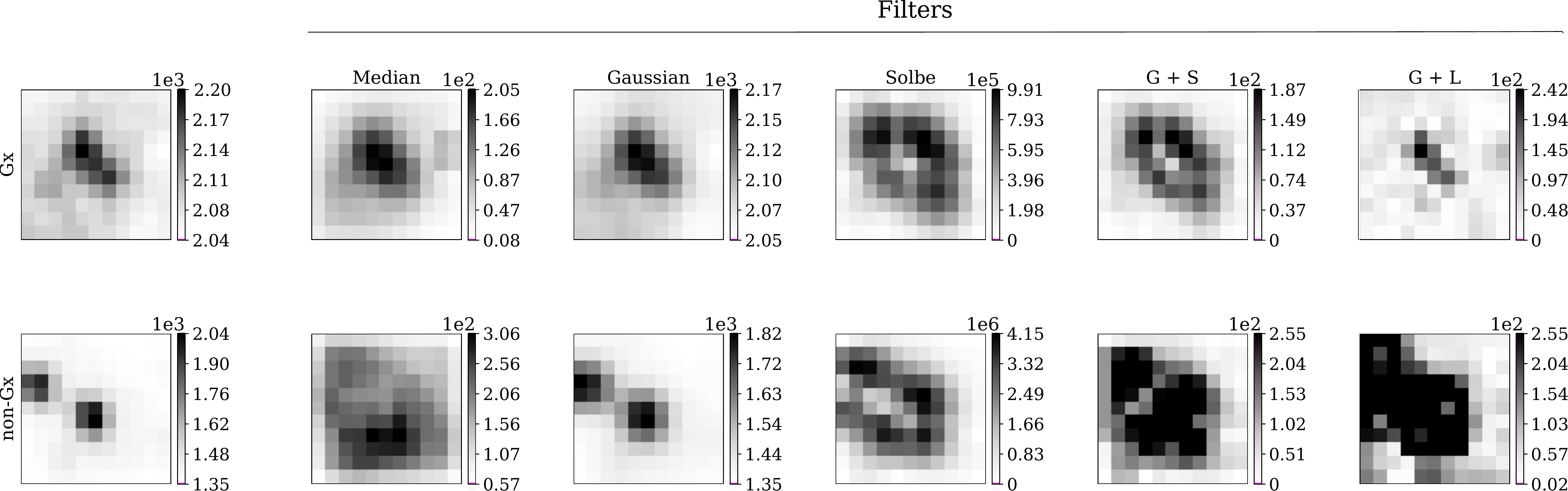} %filters_G.pdf
    \caption{J-band images with different filters. In the top row, a Gx and in the bottom row, a non-Gx.}
    \label{fig:mask}
\end{figure*}

\textbf{Filters:}
We used separate and combined smoothing and edge detection filters and analysed whether they were optimal for the separation of Gxs and non-Gxs in the centre of the images. As smoothing filters, we used the median and Gaussian filters. The median filter is a shift invariant linear filter, which replaces each pixel by a linear combination of its neighbours (which can include itself). For this experiment we used a 3$\times$3 filter with values equal to 0.8. The Gaussian filter is also a linear filter in which the values are taken from the Gaussian distribution, so that the nearest neighbouring pixels have more weight than pixels further away from the central pixel. Regarding edge filters, we used the Solbe and Laplace filters. Since the edges are discontinuities, the first one removes the noise and mimics the first derivative, i.e. it takes the difference in pixel intensities, thus detecting sharp jumps in pixel intensities. The second filter also removes the noise but in this one it performs the second derivative, thus detecting edges that have gradual changes in intensities.
Figure~\ref{fig:mask} shows an example of the changes in the image of a Gx when the different filters are applied.

\subsubsection{Selection}

In order to select the most important features, we made a visual inspection of the appearance of the images with the different sizes, passband differences, filters and scaling as well as with the images of the cluster centres resulting from the \textit{k}-means method. These centres are the average values of the objects in a cluster, thus indicating the type of objects that characterise each cluster, e.g. images with point objects or extended central objects.  In addition, for each case we analysed the segregation of Gx and non-Gx from the percentage of the total number of Gxs and non-Gxs objects in each cluster. The results are presented in  Table~\ref{tab:Performance_K_means}. The sums of columns 2 and 4 and of columns 3 and 5 are equal to 100. Therefore, the case of perfect classification occurs when each cluster contains a purely Gx or non-Gx set, i.e. cluster 0: $0$ Gx $|$ $100$ non-Gx and cluster 1: $100$ Gx $|$ $0$ non-Gx in percentage.  Finally, we examine whether the class distributions are separated into a two principal component space. 

\begin{table}
	\centering
	\caption{ Segregation of objects labelled as Gxs and non-Gxs in the two clusters resulting from \textit{k}-means when using different features of the 11 \nobreak $\times$ 11 $\times$ 3 images.}
	\label{tab:Performance_K_means}

	\begin{tabular}{l|c|c||c|c|} 
        \hline
         IS samples           & \multicolumn{4}{| c |}{Segregation}\\
                              
                              &  \multicolumn{2}{| c |}{cluster 0} & \multicolumn{2}{| c |}{cluster 1}\\
                              
                              & Gx [\%] & non-Gx [\%] & Gx [\%] & non-Gx [\%]\\
         \hline
          Original:             &         &             &         &  \\
          \hspace{0.2 cm}                    & 28      & 26          &  72     & 74\\
         \hline
          Passband differences:   &    &     &    &   \\
          \hspace{0.2 cm} $J-K_s$             & 84 & 78   & 16 & 22\\ 
          \hspace{0.2 cm} $J-H$               & 79 & 86   & 21 & 14\\
          \hspace{0.2 cm} $H-K_s$             & 71 & 84   & 29 & 16\\ 
          \hspace{0.2 cm} $H-K_s$ \& $J-K_s$   & 60 & 47   & 40 & 53\\ 
          \hline
		  Scaled:              &    &     &    &   \\
          \hspace{0.2 cm} scl\_image           & 50 & 56  & 50 & 44\\ 
          \hspace{0.2 cm} scl\_band            & 63 & 62  & 37 & 38\\ 
          \hspace{0.2 cm} scl\_band\_set       & 65 & 69  & 35 & 31\\ 
          \hspace{0.2 cm} scl\_band\_tile      & 6  & 7   & 94 & 93\\  
          \hline
          Filters:             &    &     &    &   \\
          \hspace{0.2 cm} Median              & 47  & 36  & 53 & 64\\  
          \hspace{0.2 cm} Gaussian            & 71  & 73  & 29 & 27\\  
          \hspace{0.2 cm}  G + S       & 59  & 61  & 41 & 39\\  
          \hspace{0.2 cm}  G + L       & 45  & 57  & 55 & 43\\  
          \hline
	\end{tabular}
\end{table}

\vspace{0.4 cm}

\begin{figure*}
\centering
\includegraphics[width=1\linewidth
]{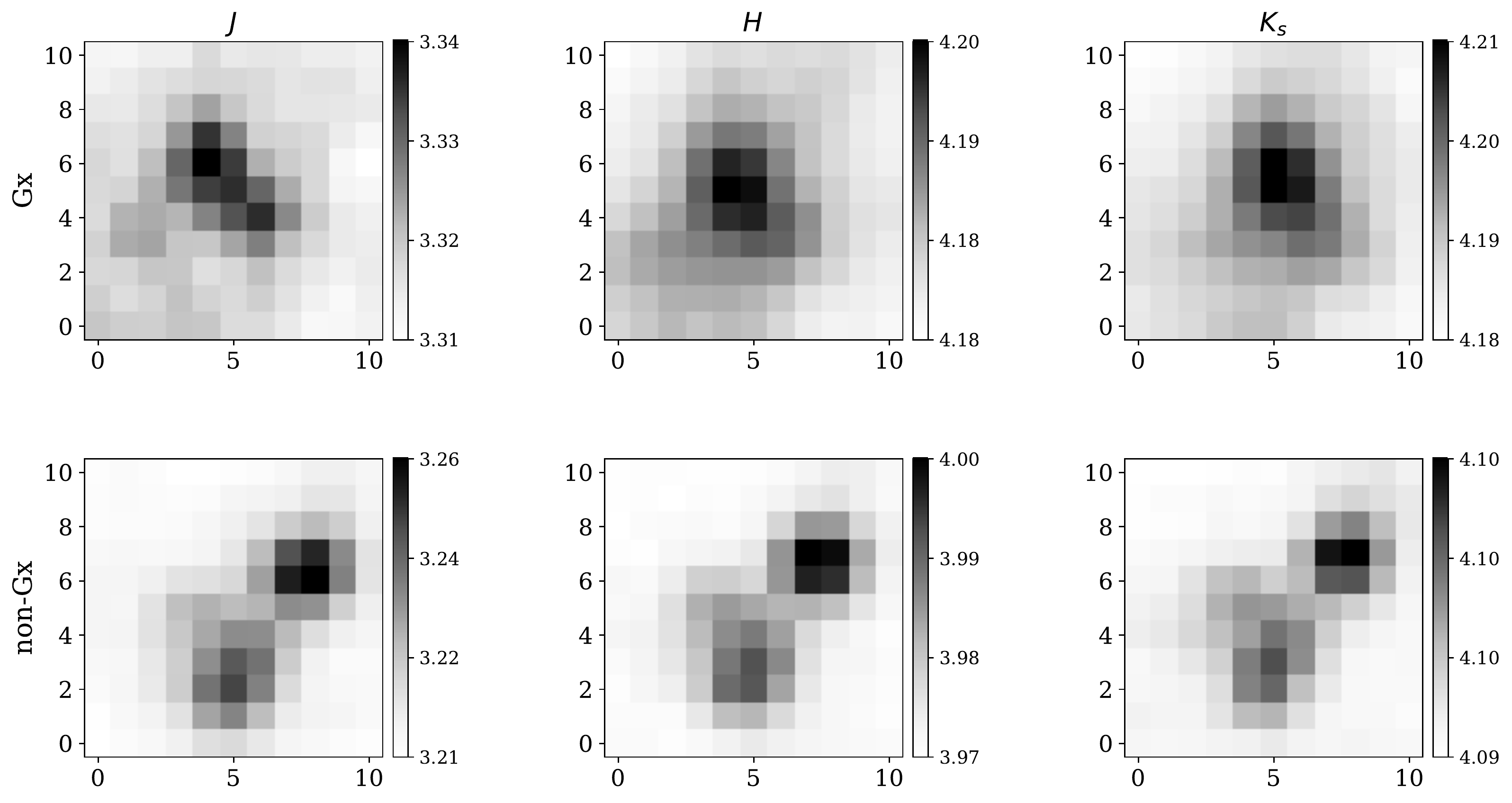}
\caption{Images of 11 $\times$ 11 pixels with examples of a Gx (upper panels) and a non-Gx (bottom panels) in the $J$, $H$ and $K_{s}$ passbands. The colour bar indicates the intensity in each pixel in logarithmic scale.}\label{fig:3filters}
\end{figure*}

\begin{figure}
    \centering
    \includegraphics[width=0.35\linewidth]{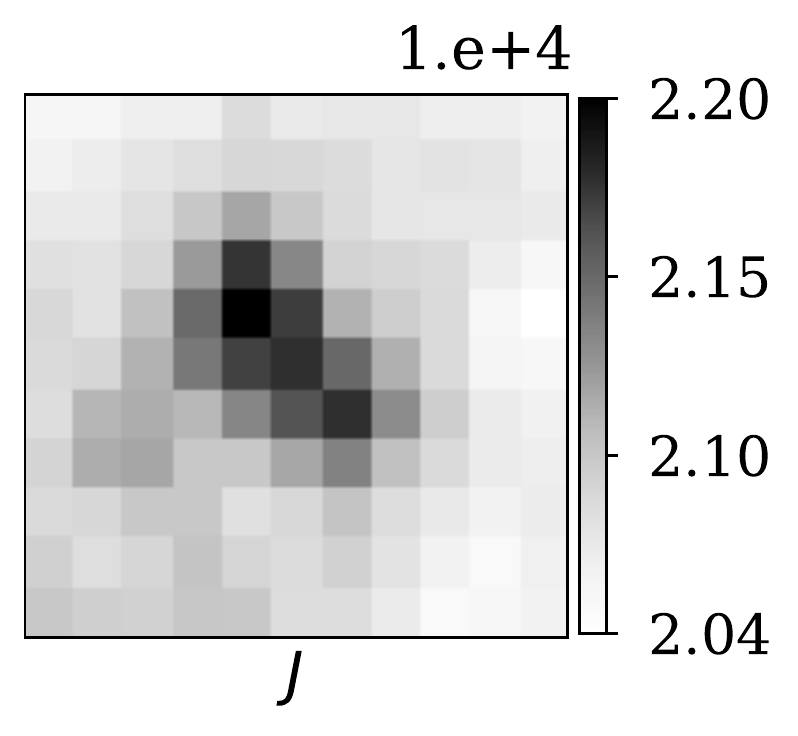}
    \hspace{0.5 cm}
    \includegraphics[width=0.35\linewidth]{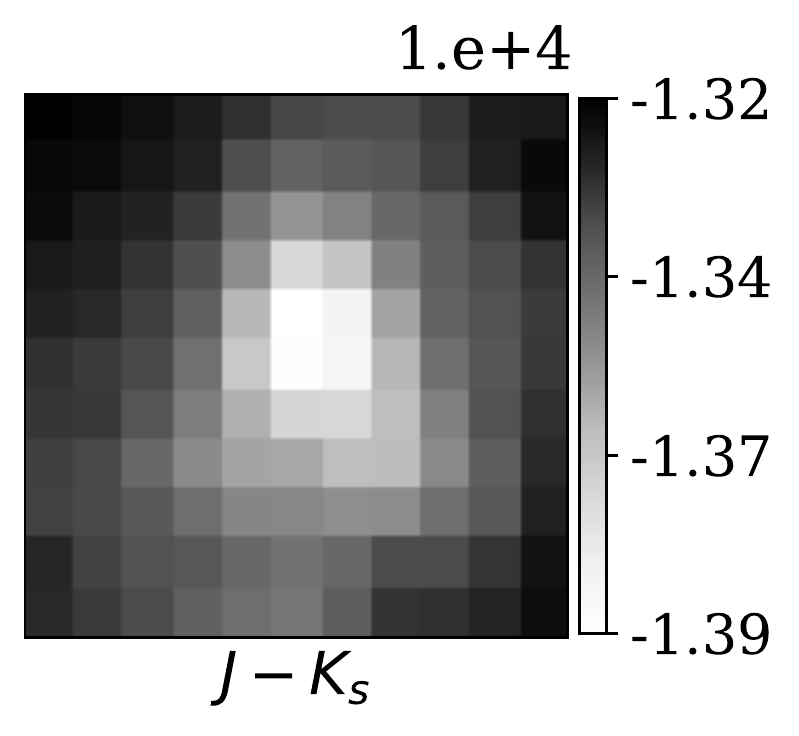}
    \includegraphics[width=0.35\linewidth]{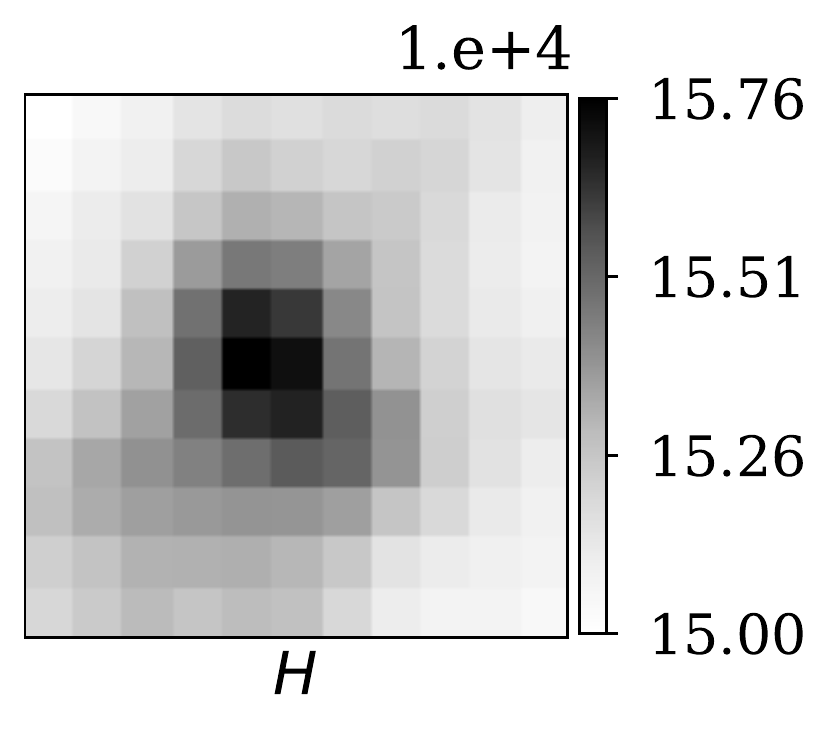}
    \hspace{0.5 cm}
    \includegraphics[width=0.35\linewidth]{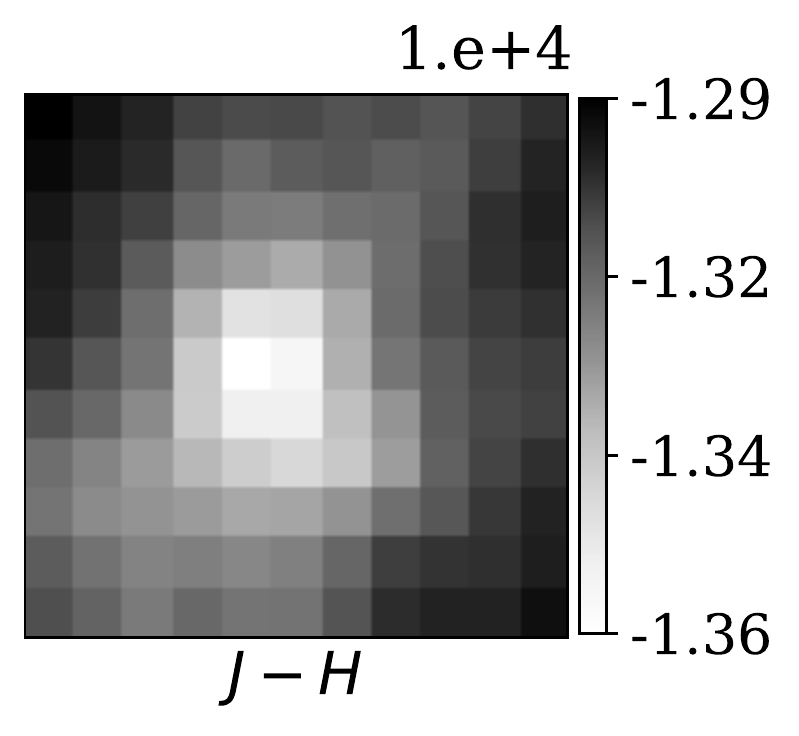}
    \includegraphics[width=0.35\linewidth]{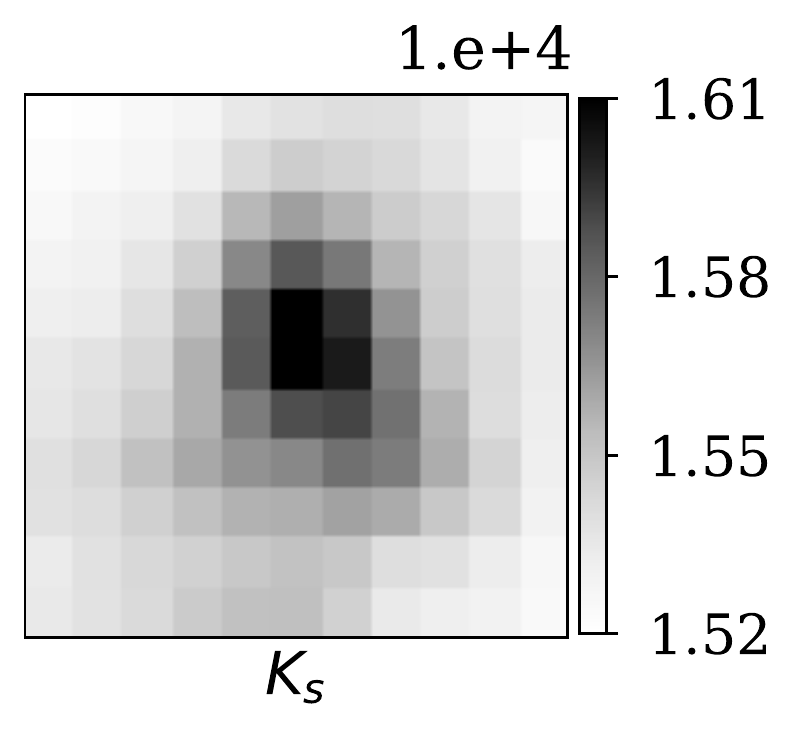}
    \hspace{0.5 cm}
    \includegraphics[width=0.35\linewidth]{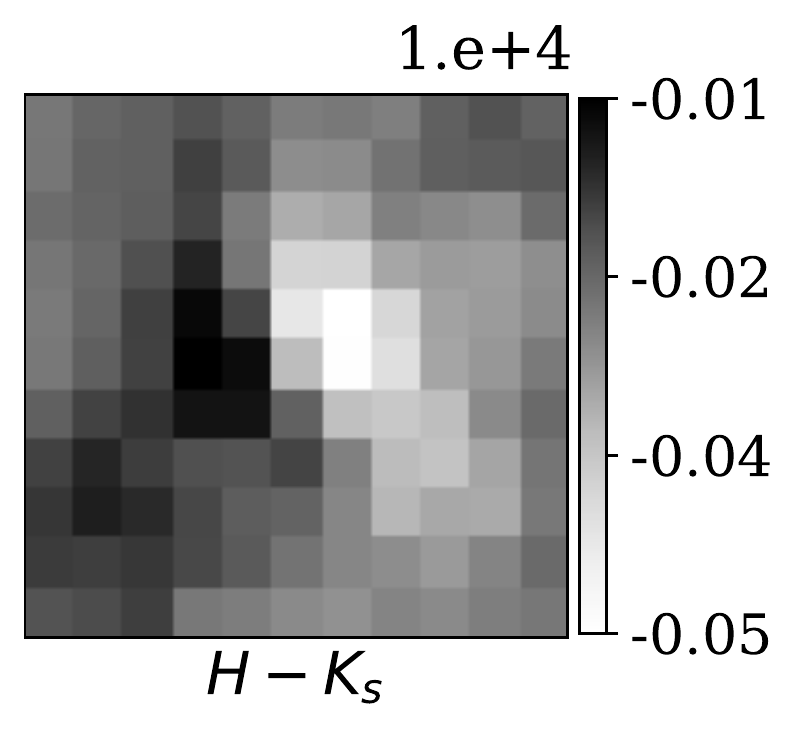}
    \caption{11 $\times $11 pixels images of a galaxy.  The left panels show the galaxy in the $J$, $H$ and $K_s$ passbands.  Right panels correspond to $J-K_s$, $J-H$ and $H-K_s$ passbands differences.}
    \label{fig:color}
\end{figure}

 The comparison of the results by statistical methods shows that the image size to be used is the 11 $\times$ 11 pixels. In Figure~\ref{fig:3filters}, we present an example using this size of images in the three passbands for a Gx and non-Gx. This size allows us to have a good sampling of the central parts of the objects minimizing stellar contamination.  It also gives a better performance in the separation between Gx and non-Gx according to the \textit{k}-means method. As for the passband difference, since the intensities of the $K_s$- and $J$-passbands have an order of magnitude difference, with the $K_s$-band intensities being higher, the resulting images are similar to those of the $K_s$-band. This is also the case when we make the difference between the $H$ and $K_s$ passbands, but not when we make the difference between $J$-$H$, in which case the shape of the central object in the image is deformed. Figure~\ref{fig:color} shows an example of these differences for a typical Gx image in the right panels together with the object in the three passbands in the left panels.  The combination of passbands differences of $J$-$K_s$ and $H$-$K_s$ improves the separation between Gxs and non-Gxs compared to using just one of these passbands differences, since we found that most Gxs and non-Gxs do not belong to the same cluster, with a balance in cluster 0 of  $60$ Gx $|$ $47$ non-Gx and cluster 1 of   $40$ Gx $|$ $53$ non-Gx (see Table~\ref{tab:Performance_K_means}). 


\begin{figure}
    \centering
    \includegraphics[width=1\linewidth]{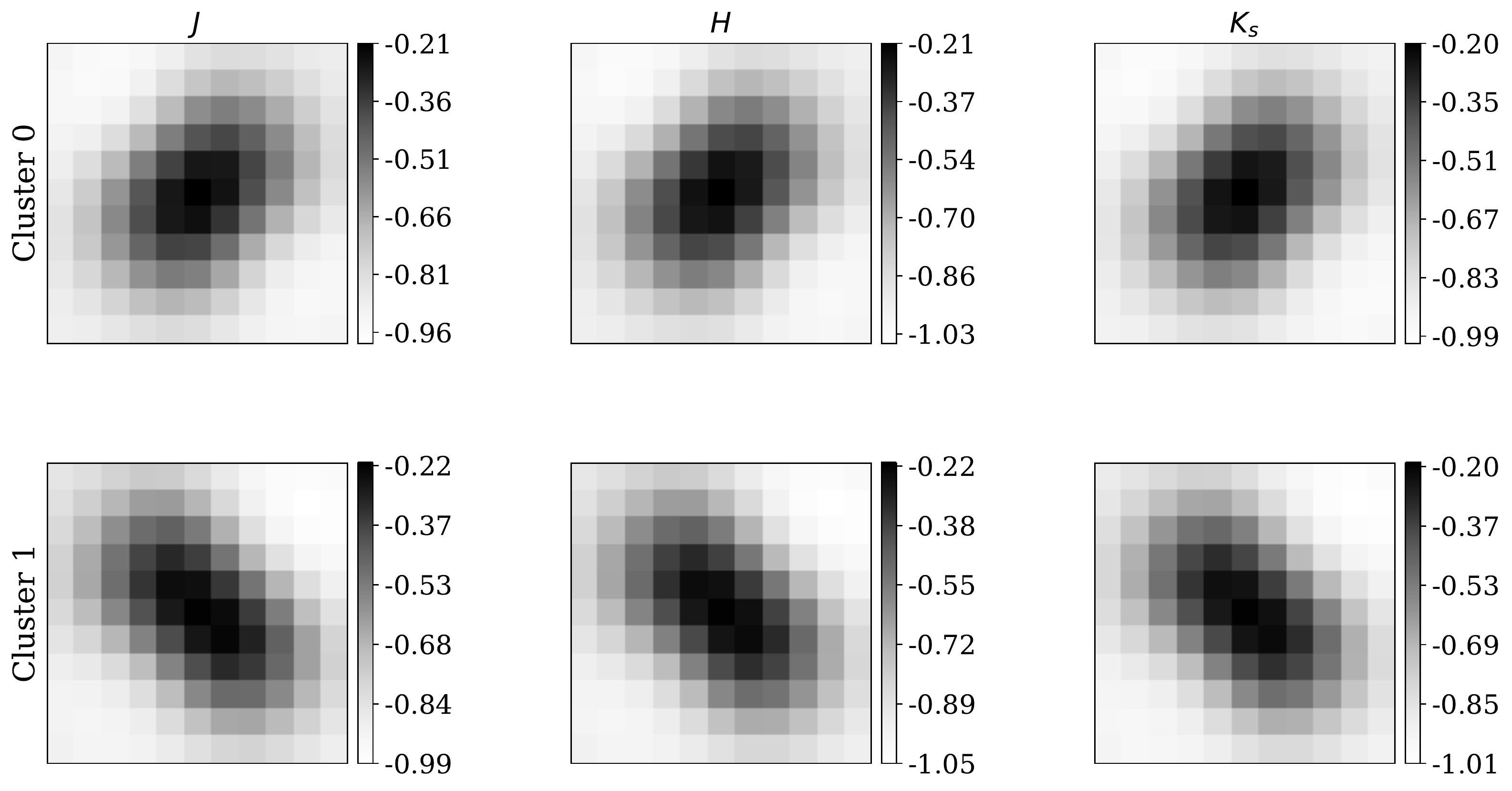}
    \caption{Images in the $J$, $H$ and $K_s$ passbands of the cluster centres obtained with the \textit{k}-means method applied on the 11 $\times$ 11 pixels images scaled with the scl\_band procedure.}
    \label{fig:cluster_centers}
\end{figure}

 As far as scaling is concerned, with the exception of the scl\_band\_tile method, all the other methods preserve the appearance of the central objects when reconstructing their images. However, this does not happen when we analyse the images corresponding to the cluster centres, where the scl\_band scaling method, in addition to distinguishing ranges of intensities (high and low), also shows the distinction of extended objects in the centre of the image with different directions of elongation as shown in Figure~\ref{fig:cluster_centers}. This information is often relevant in  supervised models such as a CNN \citep{Cheng_2021}. Regarding the purity of each class (Gx and non-Gx) in each cluster, no tendency of segregation of these classes is observed (see Table~\ref{tab:Performance_K_means}). 

\begin{figure}
    \centering
    \includegraphics[width=1\linewidth]{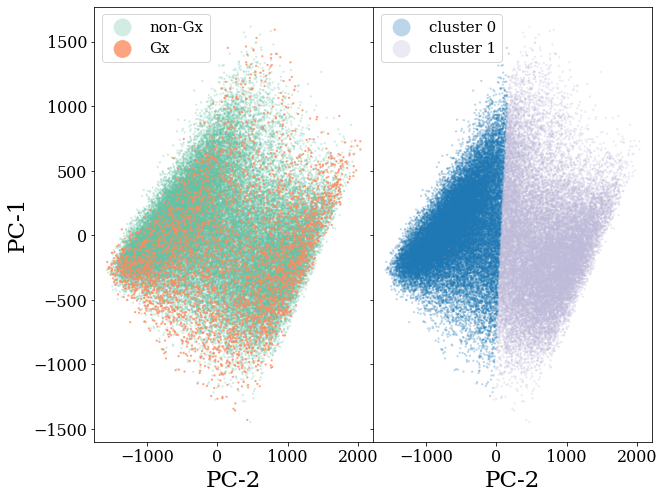}
    \caption{Distribution of objects with features generated with the combination of Gaussian and Laplace filters in a space of two principal components. {\it Left panel}: The object types Gx and non-Gx are distinguished with orange and green colours, respectively. {\it Right panel}: Objects in clusters 0 and 1 are distinguished with blue and pink colours, respectively.}
    \label{fig:dist_gl}
\end{figure}

With respect to the features generated with the filters, the images with smoothing filters do not show a significant effect in a classification from a visual inspection nor a segregation trend with the k-means method in feature space or principal component space. However, if we compare the smoothing with the median and Gaussian filters of Gx and non-Gx images, the Gaussian filter preserves the most relevant properties of the original image features, such as the shape of the galaxy. Furthermore, if we combine the Gaussian filter with an edge filter,  Gaussian+Solbe (G+S) and  on the other hand Gaussian+Laplace (G+L)  we can find a smooth structure in the object distribution that separates Gx objects from non-Gx objects in a space of two principal components. Figure~\ref{fig:dist_gl} shows the distribution of objects with features generated with the combination of Gaussian and Laplace filters in a two principal component space. We distinguish objects that are in clusters 0 and 1 as well as Gx and non-Gx. In this reduced space, most of the Gx objects are distributed on the edges with positive slope. However, the number of Gxs in each cluster is comparable. With this feature generation method, the combination of gauss+Laplace filters is better in terms of segregation of object types in each cluster, such that, cluster 0: $45$ Gx $|$ $57$ non-Gx and cluster 1: $55$ Gx $|$ $43$ non-Gx in percentage.

Based on the obtained results, we construct three sets of IS. One of them has only the features scaled with the  scl\_band  methodology and we call it $\mathrm{IS_{SCL}}$. Another set of IS, called $\mathrm{IS_{DF}}$ has also the features scaled by scl\_band plus features generated with the passband differences $J$-$K_s$ and $H$-$K_s$. The third set consists in using features scaled with scl\_band plus the features generated by applying a combined gauss+Laplace filter in the $J$, $H$ and $K_s$ passbands, we call this set $\mathrm{IS_{FLT}}$.

\subsection{Photometry-based Samples}

As for the PS sample, we also selected three different sets of features to reduce the complexity of the problem and/or to find the best performance of the models. We describe these sets in the following.

\subsubsection{All features}

One of the feature sets considered is PS described in subsection~\ref{subsec:PS}, which consists of 22 photometric and morphological features. From now on, we will refer to this set of features as $\mathrm{PS_{ALL}}$.

\subsubsection{Mutual Information}

\begin{figure}
    \centering
    \includegraphics[width=\columnwidth]{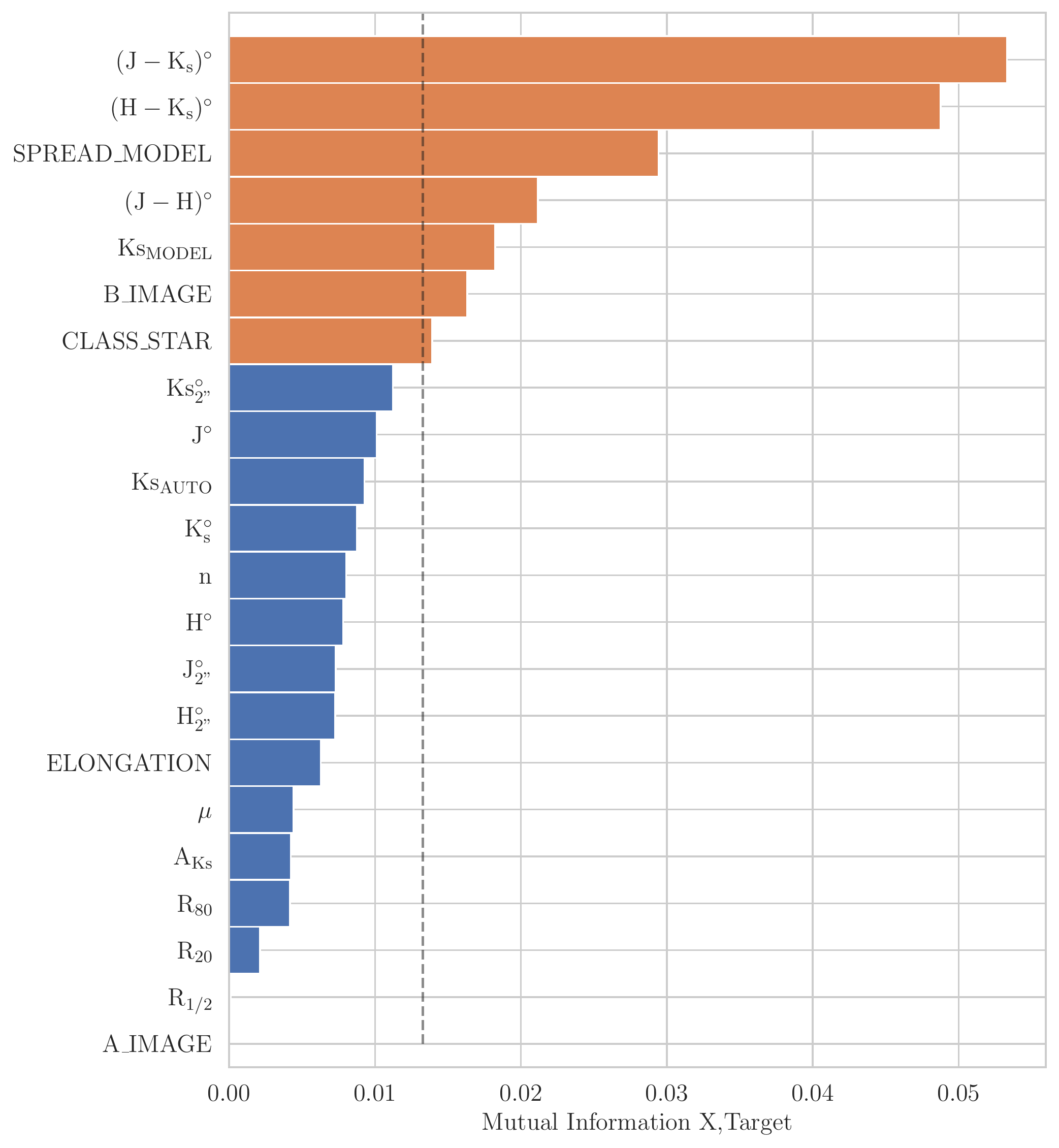}
    \caption{Scoring of features by MI. In orange, the selected features and in blue, the ignored ones.}
    \label{fig:mi_PS}
\end{figure}

 We applied the MI method on the PS  sample with the objective of measuring the information about the target variable (i.e., Gx and non-Gx classification) retrieved by observing each of the photometric and morphological parameters. Based on this analysis, we selected the seven features with the best score: $(J-K_s)_{2}^{0}$, $(H-K_s)_{2}^{0}$, SPREAD\_MODEL, $(J-H)_{2}^{0}$, $K_{s _{\textrm{MODEL}}}, B_{\textrm{IMAGE}}$ and CLASS\_STAR. Figure~\ref{fig:mi_PS} shows the order of importance of each feature according to this method with Gx and non-Gxs labels. From here on we will refer to this set as $\mathrm{PS_{MI}}$.

\subsubsection{Principal Component Analysis}

\begin{figure}
    \centering
    \includegraphics[width=\columnwidth]{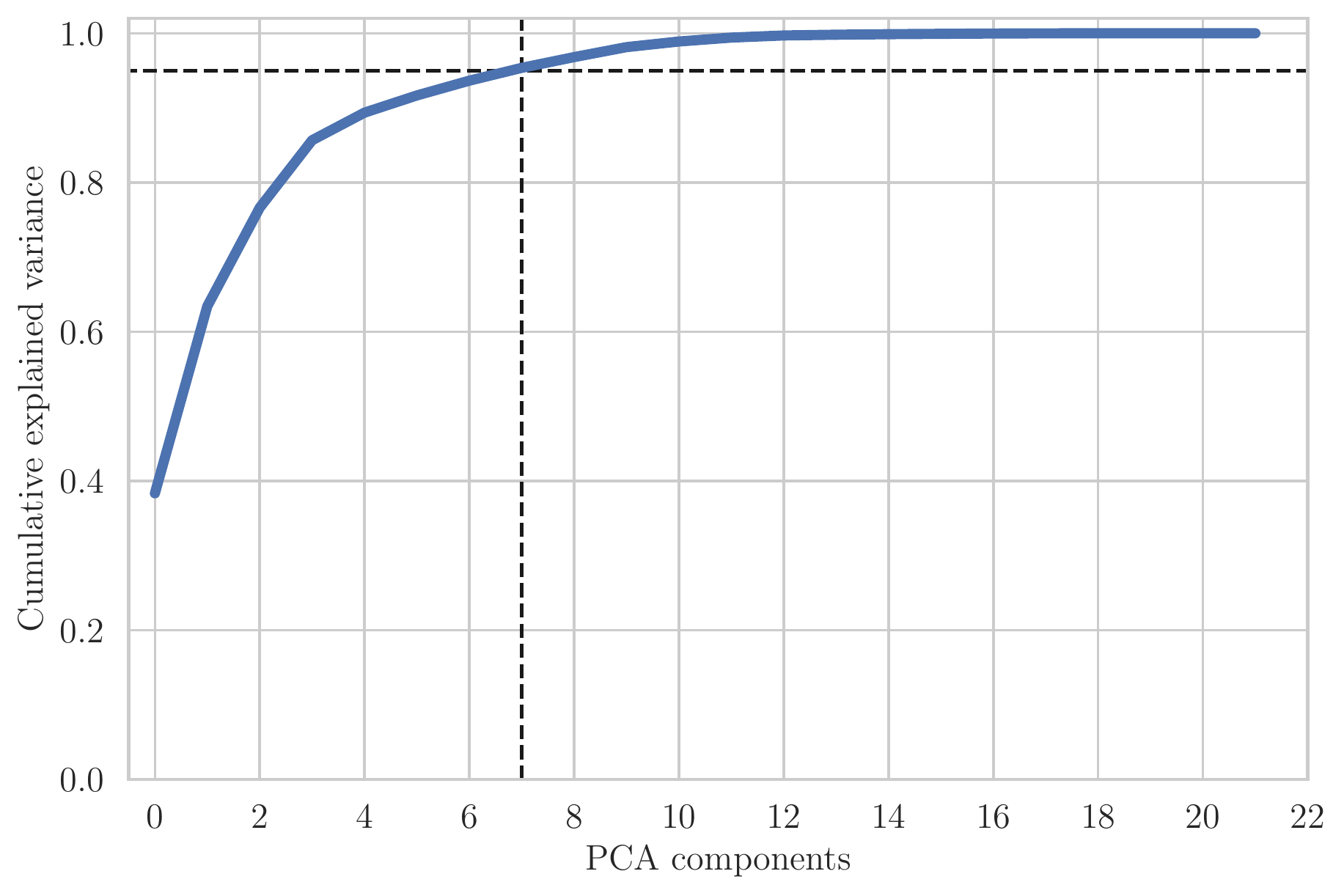}
    \caption{Cumulative variance explained from PCA applied to the photometric sample. The dashed vertical line at seven crosses the blue curve at approximately 98 per cent of the cumulative explained variance (dashed horizontal line).} 
    \label{fig:pca_PS}
\end{figure}

In addition, we performed a PCA using the total number of features contained in the PS sample, except the target. To do so, we previously standardised all features by applying a z-scaler with mean equal to zero and standard deviation equal to one. From this analysis we selected eight components representing in total 98 per cent of the cumulative explained variance, as can be seen in Figure~\ref{fig:pca_PS} where this statistic is plotted against the number of considered features. Hereafter, we will refer to this set as $\mathrm{PS_{PCA}}$.

\section{Identifying galaxies}\label{sec:Identifying_galaxies}

The relationships we find between features and object type through the statistical methods in IS and PS are useful for feature selection but insufficient for classifying Gxs and non-Gxs. Supervised learning requires a training set to learn the underlying correlations between input and targets. Since the performance of machine learning models varies with the number of features and the feature type, we studied the performance of some models when inputting data from different types of images and when inputting different photometric datasets.

\subsection{Models}

There is a large amount of bibliographic material that explains machine learning models in detail. Below we briefly describe each of the models used in this paper, and encourage the reader to examine the corresponding references for more details.

\vspace{0.4 cm}

{\it Linear support vector classification (LSVC)}: 
Support vector classification assigns categorical data from the training set points in a space. From the distribution of the points in space in the training stage, a hyperplane is defined that separates two categories. New examples are assigned a category based on which side of the separation hyperplane they are located \citep{LinearSupportVector}.

\vspace{0.2 cm}

{\it Neural networks (NN)}:
A standard NN is made up of many simple, connected processors, called neurons. In deep learning, the NN contains an input layer, one or more hidden layers and an output layer. Each artificial neuron is connected to other neurons and has an associated weight and threshold. Learning or weight assignment consists of finding the weights that cause the NN to exhibit the desired behaviour. Depending on the problem and how the neurons are connected, such behaviour may require long causal chains of computational steps, with each step typically transforming the output of the neurons in a non-linear way \citep{SchmidhuberJ_2015}. In particular, we used convolutional neural networks (CNNs) because they have proven to be extremely successful in the morphological classification of galaxies \citep{2021MNRAS.507.4425C}. For their use, we consider images in a three-dimensional space where each dimension is a passband.\\

{\it Random forest (RF)}: Random forests are a combination of decision trees such that each tree depends on the values of a random vector sampled independently and with the same distribution. The parameters of a random forest are the variables and thresholds used to split each node during training \citep{2001_Breiman}.

\vspace{0.2 cm}

{\it Gradient boosting (XGBoost)}: This machine learning technique builds a prediction model in the form of an ensemble of weak predictor models. In its standard form, these weak predictors are decision trees that are assembled sequentially with the goal of helping to classify observations that were not correctly classified by the previous trees. The final prediction of the ensemble is therefore the weighted sum of the predictions made by each decision tree. The advantage of this algorithm over other boosted methods is that it allows optimisation of an arbitrary differentiable loss function \citep{XGBoost}.

\vspace{0.2 cm}

Given the large imbalance between Gx and non-Gx examples, we analysed whether balancing the number of examples of these two objects in the training set improved the performance of some models. For IS we use the class balancing  method \textsc{SMOTETomek}, this method increases
 the class of Gxs instances via the \textsc{SMOTE} \citep[Synthetic Minority Over-sampling Technique,][]{Chawla_2002} technique, discovering Gxs-like synthetic objects in the feature space along line segments joining any/all k-nearest neighbours of class Gxs. The second technique is \textsc{Tomek links} \citep{Ivan_1976}, it generates samples from a given dataset preserving or improving the performance that would be obtained in the classification. For  PS we implement the RF balancing algorithm developed by \citet{ImbalancedLearn}, which considers this imbalance when training the model. Specifically, each tree in the forest is trained using a balanced subsample that is constructed by subsampling the majority class (i.e., non-galactic objects)

\subsection{Metrics}

In this paper, and for the purpose of specifying the evaluation metrics, we will refer to non-Gxs objects as the negative class, while the positive class are Gxs objects. Taking this into account, True Negative (TN) and True Positive (TP) objects are those correctly classified as non-Gx and Gx, respectively. On the other hand, False Negative (FN) objects are Gxs classified as non-Gx objects (type II error); and False Positive (FP) objects represent those objects incorrectly classified as Gxs (type I error). These four values are summarised in the confusion matrix as shown in Table~\ref{tab:ConfusionMatrix}. For an ideal classifier, the non-diagonal elements of this matrix should be equal to zero.

\begin{table}
    \centering
    \caption{Confusion matrix}
    \label{tab:ConfusionMatrix}
    \begin{tabular}{|c|c|c|}
        \hline
        & \multicolumn{2}{|c|}{Predicted} \\
        \hline
        \multirow{2}{*}{\rotatebox[origin=c]{90}{Actual}} & TN & FP \\
        & FN & TP \\
        \hline
    \end{tabular}
\end{table}

The metric that is commonly used in classification problems is the accuracy metric, which is the ratio of the number of correct predictions to the total number of input samples. However, reporting the performance of models with this metric in an unbalanced classification problem such as our case is not the best option. Therefore, to compare the performance of the models we use four metrics, the aforementioned confusion matrix, recall, precision and F1 score. The last three metrics are calculated for the positive cases, i.e., the Gxs.

With the precision metric we can measure the quality of the models to correctly classify objects, it answers the question of what percentage of objects classified as Gxs are Gxs, and for  calculation we use the values of the right column of the confusion matrix (Table~\ref{tab:ConfusionMatrix}) in equation~\ref{equ:precision}

\begin{equation}
    \hspace{2.8 cm} \textit{precision} =  \frac{TP}{TP + FP}
    \label{equ:precision}
\end{equation}

The recall metric is a measure of the fraction of galaxies that are identified as such by the model. This measure can be computed through the values in the last row of the confusion matrix, making use of equation~\ref{equ:recall}.

\begin{equation}
    \hspace{2.8 cm} \textit{recall} =  \frac{TP}{TP + FN}
    \label{equ:recall}
\end{equation}

It is common, in unbalanced binary classification problems, to illustrate recall and precision metrics for different decision thresholds, i.e. different discrimination values at which we decide that a case is positive according to their probability. These illustrations are known as Recall-Precision curves and show the binary classifier ability according to a threshold. The most commonly used value for this threshold is 0.5, which we have used to calculate the values of TP, TN, FP and FN and, consequently, of all the metrics mentioned above. 

The F1 is a combination of precision and recall calculated through their harmonic mean (equation~\ref{equ:f1}). Therefore, the resulting values of this quantity consider both the quality and the quantity of the Gxs classifications made by the models, so we have used it to determine which is the best classifier of Gxs and non-Gxs trained with IS and PS.

\begin{equation}
    \hspace{3 cm} F1 = 2\, \frac{\textit{precision} \cdot \textit{recall}}{\textit{precision} + \textit{recall}}
    \label{equ:f1}
\end{equation}

\subsection{Classification}\label{sub:identification}

In the Tables~\ref{tab:performanceIS} and ~\ref{tab:performancePS} we show the performance obtained with each of the models on the test set with information of IS and PS, respectively. We chose as best models those with the highest F1 score calculated for the objects of type Gxs. In the case of IS, the best model was the CNN trained with an unbalanced training set of 44 $\times$ 44 pixels images with six channels, the first three correspond to the passbands $J$, $H$ y, $K_s$ and the last three correspond to the edges detected with the Gaussian and Laplace filters implemented on the three passbands of an image,  i.e, the  $\mathrm{IS_{FLT}}$.  Regarding PS, the best model was XGBoost trained with 22 photometric and morphological features, i.e, the $\mathrm{PS_{ALL}}$ sample. The detailed description of the models is in Appendix~\ref{app:TheBetsModels}.

\begin{table}
\centering
\caption{IS performance on the testing set measured with precision, recall and F1 metrics of the objects Gx. The $^*$ symbol indicates that the training set is out of balance. The model with the highest F1 value is highlighted.} \label{tab:performanceIS}
\begin{tabular}{crccc}
\hline  
 Sample & Model & Precision & Recall & F1\\
\hline
\multirow{4}{*} {\rotatebox[origin=c]{90}{\hspace{0.5 cm }$\mathrm{IS_{SCL}}$}}
  & LSVC  & 0.19 & 0.52 & 0.27\\
  & CNN*  & 0.80 & 0.58 & 0.67\\
  & RF    & 0.45 & 0.61 & 0.52\\
\hline  

\multirow{4}{*} {\rotatebox[origin=c]{90}{\hspace{0.5 cm }$\mathrm{IS_{DF}}$}} 
  & LSVC & 0.18 & 0.51 & 0.27\\
  & CNN  & 0.42 & 0.79 & 0.55\\
  & RF   & 0.46 & 0.64 & 0.54\\

\hline
\multirow{4}{*} {\rotatebox[origin=c]{90}{\hspace{0.5 cm }$\mathrm{IS_{FLT}}$}}
  & LSVC & 0.19 & 0.52 & 0.27\\
  & {\bf CNN$^*$} & {\bf 0.73} & {\bf 0.63} & {\bf 0.68}\\
  & RF & 0.44 & 0.59 & 0.51 \\
  \hline
\end{tabular}
\end{table}

\begin{table}
\centering
\caption{PS performance on the testing set measured with the precision, recall and F1 metrics of the objects Gx. The $^\dagger$ symbol indicates that the dataset is balanced. The model with the highest F1 value is highlighted.}
\label{tab:performancePS}
\begin{tabular}{crccc}
\hline  
 Sample & Model & Precision & Recall & F1\\
\hline
\multirow{4}{*} {\rotatebox[origin=l]{90}{$\mathrm{PS_{ALL}}$}}  
  & RF$^\dagger$ & 0.37 & 0.75 & 0.50 \\
  & NN & 0.78 & 0.56 & 0.64 \\
  & {\bf XGBoost} & {\bf 0.64} & {\bf 0.65} & {\bf 0.64}\\
\hline  

\multirow{4}{*} {\rotatebox[origin=l]{90}{$\mathrm{PS_{MI}}$}}  
  & RF$^\dagger$ & 0.35 & 0.74 & 0.48 \\
  & NN & 0.75 & 0.44 & 0.55 \\
  & XGBoost & 0.60 & 0.60 & 0.60 \\
  
\hline
\multirow{4}{*} {\rotatebox[origin=l]{90}{$\mathrm{PS_{PCA}}$}} 
  & RF$^\dagger$ & 0.34 & 0.75 & 0.47 \\
  & NN & 0.75 & 0.39 & 0.51 \\
  & XGBoost & 0.60 & 0.59 & 0.59 \\
  \hline
\end{tabular}
\end{table}

\begin{figure}
    \centering
    \includegraphics[width=0.48\textwidth]{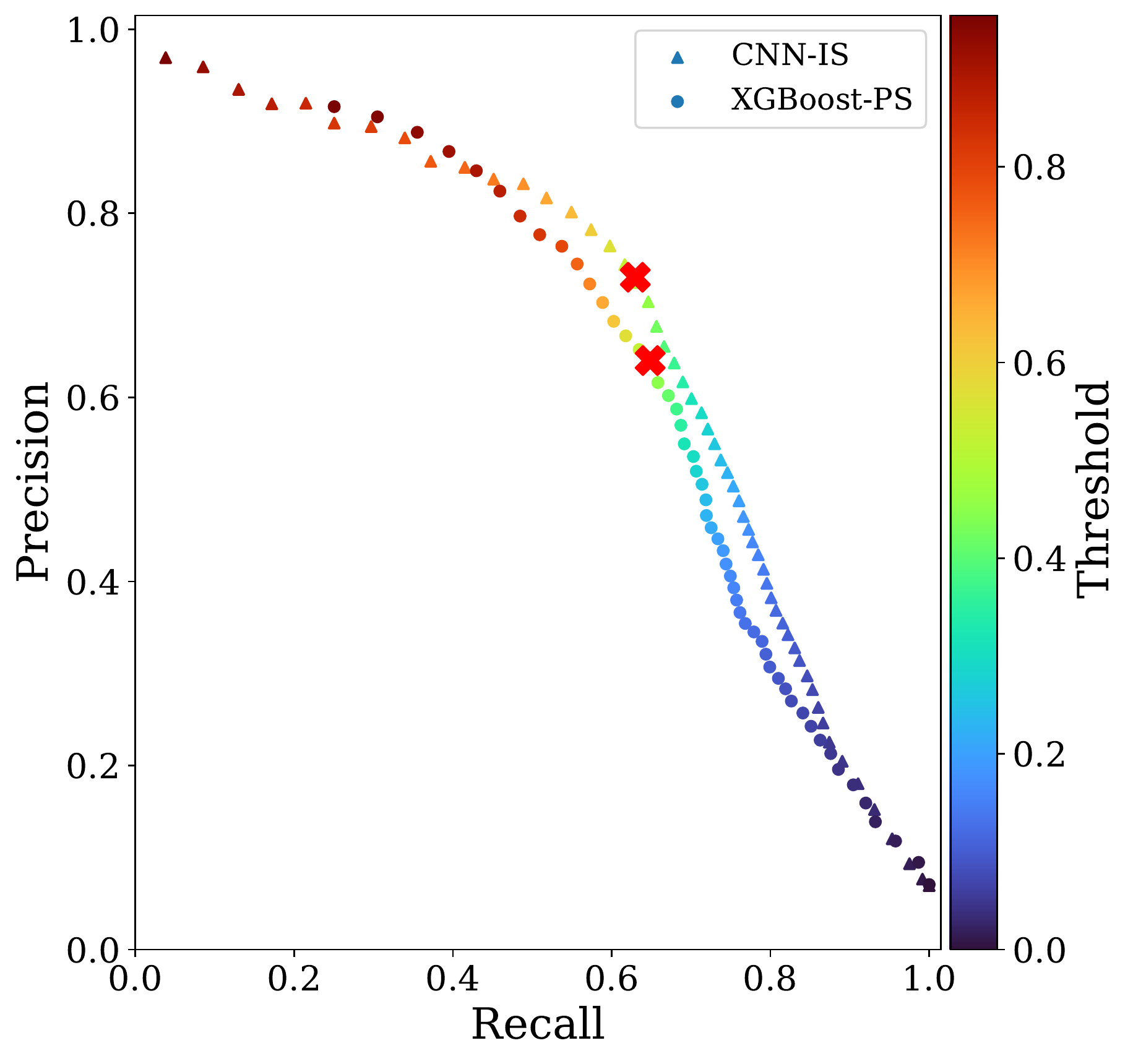}
    \caption{Recall-Precision curves for CNN and XGBoost models. The colour bar indicates the threshold values and the red cross marks indicate the threshold $= 0.5$ in each model.}
    \label{fig:recall_precision}
\end{figure}

From the outputs of the models with the best classification tested with IS and PS we obtain the probabilities of the objects to be galaxies.  with the probabilities of the candidates we created the Recall-Precision curves for different thresholds (see Figure~\ref{fig:recall_precision}). These curves can be used to generate catalogues with the lowest contamination (precision $\approx$ 1) or highest completeness (recall $\approx$ 1) desired through the classifications made by CNN, the best model obtained with IS or with the classifications made with XGBoost, the best model obtained with PS.

\begin{figure}
    \centering
    \includegraphics[width=0.45\textwidth]{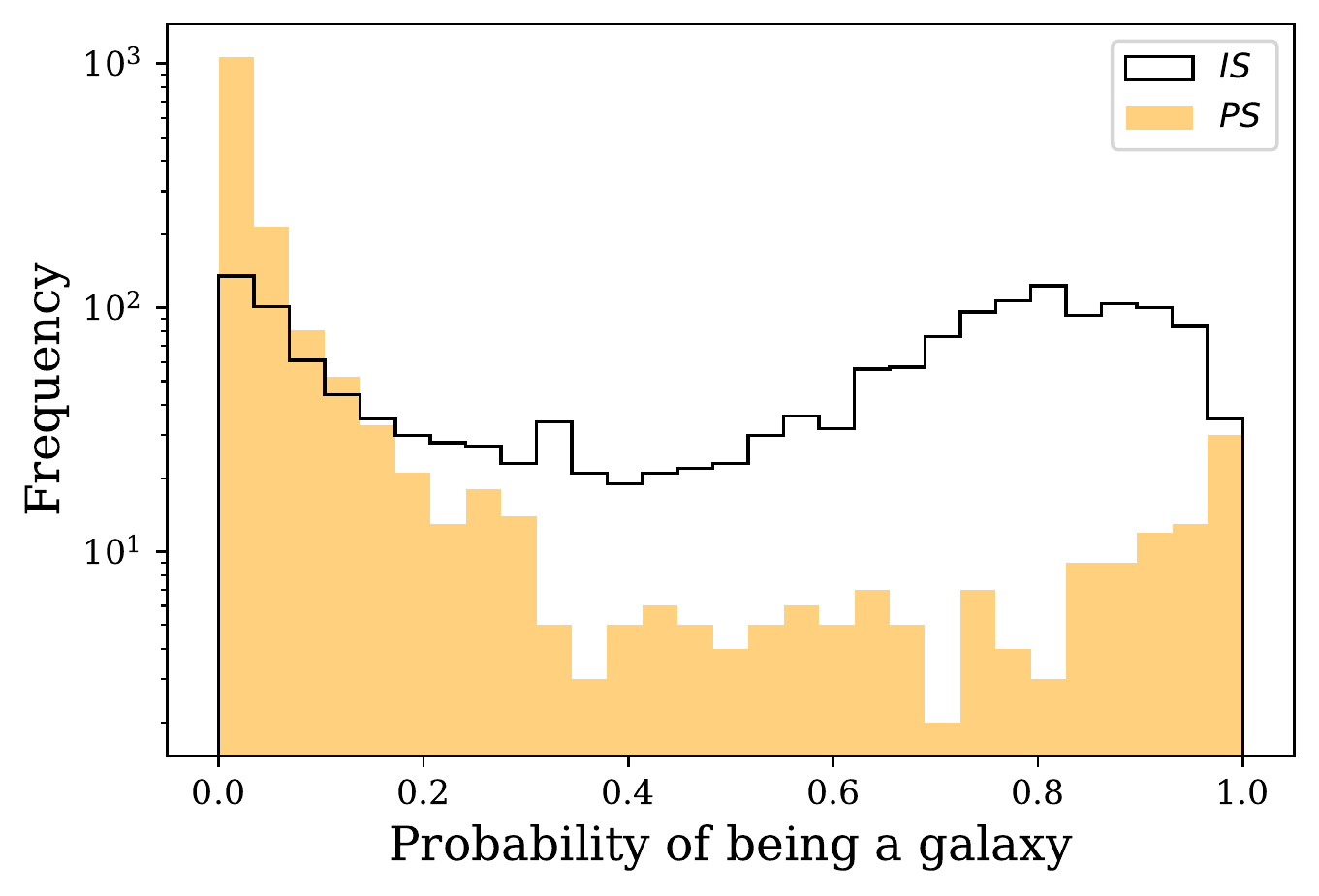}
    \caption{Probability distribution of the galaxies in the test set. In black, the probabilities of being a galaxy for the CNN model trained with IS and in orange, the distribution of probabilities of being a galaxy for the XGBoost model trained with PS. The y-axis is shown in logarithmic scale.}
    \label{fig:IP_SP_VVV}
\end{figure}

We study the assignment of probabilities that the models give to objects labelled Gx. For this, we select the objects labelled as Gx from the test set and compare the probability distributions of the CNN and XGBoost models  (see Figure~\ref{fig:IP_SP_VVV}). We find that, for probabilities greater than 0.5, the XGBoost model tends to assign probabilities close to 0.9 with high frequency compared to the CNN probability distribution, which tends to give lower probabilities to the galaxies.  
%We believe that this difference is due to the fact that the PS-trained XGBoost model uses a wide range of values for each feature while the CNN tends to use a smaller range, possibly resulting in an overfitting. 
Furthermore,  we find that the behaviour of the two distributions at low probabilities is similar, corresponding to galaxies with closer to zero SPREAD\_MODEL values.

In addition, we find a good balance between precision and recall when selecting possible extragalactic sources with probabilities greater than 0.6 for either model.  The precision and recall metrics for 1,682 candidates to be galaxies in the test set under these conditions are 0.65 and 0.69, respectively.  Therefore, this is the methodology we use to generate the galaxy catalogue in the Northern part of the Galactic disc.

\section{The VVV near-IR galaxy catalogue: Northern part of the Galactic disc}\label{sec:catalogue}

We applied the improved pipeline to the 56 tiles belonging to the Northern Galactic disc obtaining a total of 172,396 candidates to extragalactic sources. This region contains only a few sources already reported as extragalactic sources in the literature. Two 2MASS objects: DSH J1827.0-2031 \citep{2012yCat.2311....0C} and ZOA J180953.827-123353.78 \citep{2014MNRAS.443...41W} have identifications in our catalogue, they are VVVX-J182701.02-203158.3 and VVVX-J180953.85-123354.3, respectively. We visually inspected two 2MASS sources that were not found in our catalogue: J18261431-1334481 \citep{2020yCat.1350....0G} is a star and J1802473-145454 is a 2MASX object classified as Seyfert I \citep{2004ATel..246....1C} in the literature. These sources were identified with our procedure as a point sources and were rejected from our criteria to select extended sources. In the disc+20 region there are 13 HIZOA \citep{2016AJ....151...52S, 2016MNRAS.457.2366S}  and 152 LEDA \citep{1996PASJ...48..679R} objects without counterparts in our catalogue. In the NIR passbans, the HIZOA objects are extremely faint or invisible sources. By visual inspection, we detected that most of the LEDA objects are stars or stellar associations.

The candidate sources obtained with the improved pipeline have images and astrometric, photometric and morphological data. We compare some of the photometric and morphological properties of galaxies in the VVV disc and Northern disc regions. In each region we randomly take 79,700 candidates to be galaxies.  The medians of the intensities in the $J$, $H$ and $K_s$  of images in the VVV disc and Northern disc are presented in the Table~\ref{tab:medians_images_79700}. While in Table~\ref{tab:properties} we show the medians of the properties with their respective confidence intervals. 

\begin{table*}
    \caption{ Comparison of the median intensities in the J, H and K bands of the candidates belonging to the VVV and Northern disc regions in the first two columns (2) and (3), respectively, and of the galaxies in columns (4) and (5), respectively. The error is the confidence interval with a confidence level of 0.95. }
    \centering
    \begin{tabular}{lcc c cc}
    \hline
                     & \multicolumn{2}{ |c| }{Candidates}              &   \multicolumn{2}{ |c| }{Galaxies} \\
    Bands     [ADU]  & Median              & Median                        & Median   & Median\\
                     & VVV disc            & Northern disc                 & VVV disc & Northern disc\\
    \hline
     $J$             & 1561.0 $\pm$ 0.1     & 1236.8 $\pm$ 0.1         &  1542.0    $\pm$ 0.1     & 1198.8 $\pm$ 0.9\\
     $H$             & 8611.0 $\pm$ 0.4     & 4303.0 $\pm$ 0.3         &  8775.3    $\pm$ 4.4     & 4284.0 $\pm$ 3.4\\
     $K_{s}$         & 12043.1 $\pm$ 0.3    & 5109.0 $\pm$ 0.1         &  12100.5   $\pm$ 3.7     & 4980.5 $\pm$ 0.6\\
    \hline
    \end{tabular}
    \label{tab:medians_images_79700}
\end{table*}

\begin{table}
    \caption{Comparison of median photometric and morphological properties in VVV disc and Northern disc galaxies. The error is the confidence interval with a confidence level of 0.95.}
    \centering
    \begin{tabular}{lcc}
    \hline
    Properties &  Median & Median \\
    &    VVV disc   & Northern disc \\        
    \hline
     $J^{0}$ [mag]               & 16.39 $\pm$ 0.01               & 16.71 $\pm$ 0.01\\   %MAG_PSF_J_CORR
     $H^{0}$  [mag]              & 15.89 $\pm$ 0.04               & 16.22 $\pm$ 0.01\\   %MAG_PSF_H_CORR
     $K_{s}^{0}$ [mag]           & 15.77 $\pm$ 0.01               & 16.12 $\pm$ 0.12\\   %MAG_PSF_KS_CORR
     $(J-K_s)_{2}^{0}$  [mag]    & 0.63 $\pm$ 0.01                & 0.64 $\pm$ 0.01 \\   %colorJKs
     $(J-H)_{2}^{0}$    [mag]    & 0.51 $\pm$ 0.01                & 0.50 $\pm$ 0.01\\    %colorJH
     $(H-K_s)_{2}^{0}$  [mag]    & 0.12 $\pm$ 0.01                & 0.15 $\pm$ 0.01\\    %colorHKs
     $R_{20}$   [arcsec]         & 1.74 $\pm$ 0.01                & 1.71 $\pm$ 0.01\\    %FLUX_RADIUS_0 
     $R_{1/2}$  [arcsec]         & 3.35 $\pm$ 0.01                & 3.29 $\pm$ 0.01\\    %FLUX_RADIUS_1
     $R_{80}$   [arcsec]         & 6.51 $\pm$ 0.02                & 6.12 $\pm$ 0.02\\    %FLUX_RADIUS_2
     $\mu$  [mag/arcsec$^{2}$]   & 16.53 $\pm$ 0.01               & 16.55 $\pm$ 0.01\\
     $e$                         & 1.77 $\pm$ 0.01                & 1.84 $\pm$ 0.01\\
     $n$                         & 3.96 $\pm$ 0.02                & 3.83 $\pm$ 0.01\\    %SPHEROID_SERSICN
     $A_{\textrm{IMAGE}}$        & 3.43 $\pm$ 0.01                & 3.15 $\pm$ 0.01\\
     $B_{\textrm{IMAGE}}$        & 1.96 $\pm$ 0.01                & 1.72 $\pm$ 0.01\\
     A$_{Ks}$ [mag]              & 0.65 $\pm$ 0.01                & 0.44 $\pm$ 0.01\\
    \hline
    \end{tabular}
    \label{tab:properties}
\end{table}

 The medians using VVV disc data are statistically different from those obtained in the Northern disc regions both using image data and photometric and morphological data. These differences could be due to several reasons, the statistical difference of the median intensities in the images in each band, possibly caused by the observing conditions between one survey and the other. Finding a trend of one type of galaxy population caused by the difference in latitudes, as the Northern disc reaches higher and lower latitudes than the VVV disc where the interstellar extinction should be quite different. Another reason for the difference may be due to different intrinsic structures behind the regions studied by the VVV and VVVX surveys.

Using the information from these galaxy candidates and the CNN and XGBoost models trained with IS and PS, respectively, we obtained the probabilities that an object is a galaxy according to these two models. 
This list of probabilities of possible extragalactic sources can be used to generate catalogues according to the case study to be performed. Candidates are available in electronic format \footnote{\href{https://catalogs.oac.uncor.edu/vvvx_nirgc/}{NorthernGalacticDisc\_candidates.csv}}, and contain information on the equatorial coordinates, the VVVX tile to which they belong, the photometric and morphological features of $\mathrm{PS_{ALL}}$, the CNN-IS model probabilities (prob$\_$IS), the XGBoost-PS model probabilities (prob$\_$PS) and a flag indicating whether the object has visual classification, Gx $= 1$, non-Gx $= 0$ or, no inspection $= 99$. In Table~\ref{tab:candidates} we show the values of the first five possible extragalactic sources, in the electronic version we show this same information for the 172,396 extragalactic candidates.

\begin{table*}
\caption{Properties of five possible extragalactic sources obtained with the improved pipeline. The full table is available online.}
\begin{adjustbox}{angle=90}
\begin{tabular}{lcccccccccccccc}
\hline
ID [VVVX-] &   TILE &  $\alpha$ (J2000)  &  $\delta$ (J2000)  &  $K_s^{\textrm{AUTO}}$ &  $K_s^{\textrm{MODEL}}$ &  $J^0$  &  $H^0$  & $K_s^0$ &  $J_2^0$ & $H_2^0$ &  $K_s{}_2^0$ &  ($J$ -$K_{s}$)$_{2}^{0}$  &   ($J$ - $H$)$_{2}^{0}$ &  ($H$ - $K_{s}$)$_{2}^{0}$\\

 &   &  [$^{\circ}$]  &  [$^{\circ}$]  & [mag] &  [mag] &  [mag]  &  [mag]  &[mag] & [mag] & [mag] &  [mag] &  [mag]  &  [mag] &  [mag]\\
\hline
J175128.40-180341.3 &  e0982 &   267.87 &   -18.07 &  15.85 &  15.82 & 17.21 &       16.64 & 16.70 &  16.87 &   16.31 & 16.28 &  0.59 &  0.56 &  0.03\\

J175130.45-180420.8 &  e0982 &   267.88 &   -18.07 &  15.52 &  16.36 &  16.91 &  16.53 &  16.45 & 17.02 & 16.58 &  16.45 &  0.58 &  0.44 &  0.13\\

J175131.48-180335.7 &  e0982 &   267.88 &   -18.06 &  14.93 &  15.22 &  16.08 &  15.80 &  15.64 &  16.04 &  15.67 &   15.41 &  0.63 &  0.37 &  0.26\\

J175132.33-180311.9 &  e0982 &   267.88 &   -18.05 &  15.30 &  15.07 &  16.37 & 16.00 &  15.96 &  16.35 & 15.92 &  15.82 &  0.53 &  0.43 &  0.09\\

J175132.57-180417.5 &  e0982 &   267.88 &   -18.07 &  15.50 &  15.50 &  16.91 &  16.25 &  16.08 & 16.92 & 16.21 & 15.95 &  0.97 &  0.71 &  0.26\\
\hline
&&&&&&&&&&&&&&\\
\hline
&  $R_{20}$ &  $R_{1/2}$ &  $R_{80}$ & $\mu$ &  $e$  &  $n$ & $A_{\textrm{IMAGE}}$  & $B_{\textrm{IMAGE}}$  &  SPREAD &  CLASS &  A$_{Ks}$ &  prob\_IS &  prob\_PS &  visual\\

&   [arcsec] &  [arcsec] &  [arcsec] & $\left[ \frac{\textrm{mag}}{\textrm{arcsec}^2} \right]$ &   & &  &     & MODEL & STAR  & & &  &  \\
\hline
 &       1.75 &       3.08 &       6.33 &  16.53 &    1.75 & 4.28 &  2.70 &  1.54 &      0.01 &    0.29 &  0.29 &     0.01 &     0.00 & 99 \\
 
  &      2.18    &    4.41 &       11.76 &  16.70 &   2.25 & 4.89 &  3.76 &  1.67 &      0.00 &    0.01 &  0.29 &    0.02 &       0.00 & 99\\
  
 &      1.71 &        3.07 &       5.41 &  15.85 & 1.71 &  1.79 &  3.22 &  1.88 &  0.01 &    0.03 &  0.30 &  0.03 &  0.01 & 99 \\
 
 &      1.67 &        3.15 &       4.90 &  16.43 & 1.78 &  4.21 &  3.24 &  1.81 & 0.01 &     0.02 &  0.30 & 0.04 &  0.02 & 99\\
 
 & 1.33 &  3.07 &  5.98 &  16.42& 1.64 & 6.97 &  2.93 &  1.79 &  0.01 & 0.03 &  0.29 &  
 0.07 & 0.14 & 99\\
\end{tabular}
\label{tab:candidates}
\end{adjustbox}
\end{table*}

Customised galaxy catalogues in completeness and purity can be generated through the probabilities and the precision and recall of the CNN and XGBoost models. Figure~\ref{fig:recall_precision} summarises an estimate of the precision and recall values obtained by varying the decisive threshold for defining a possible extragalactic source as Gx.

\begin{figure*}
\centering
%trim=left bottom right top
\includegraphics[width=1\textwidth]{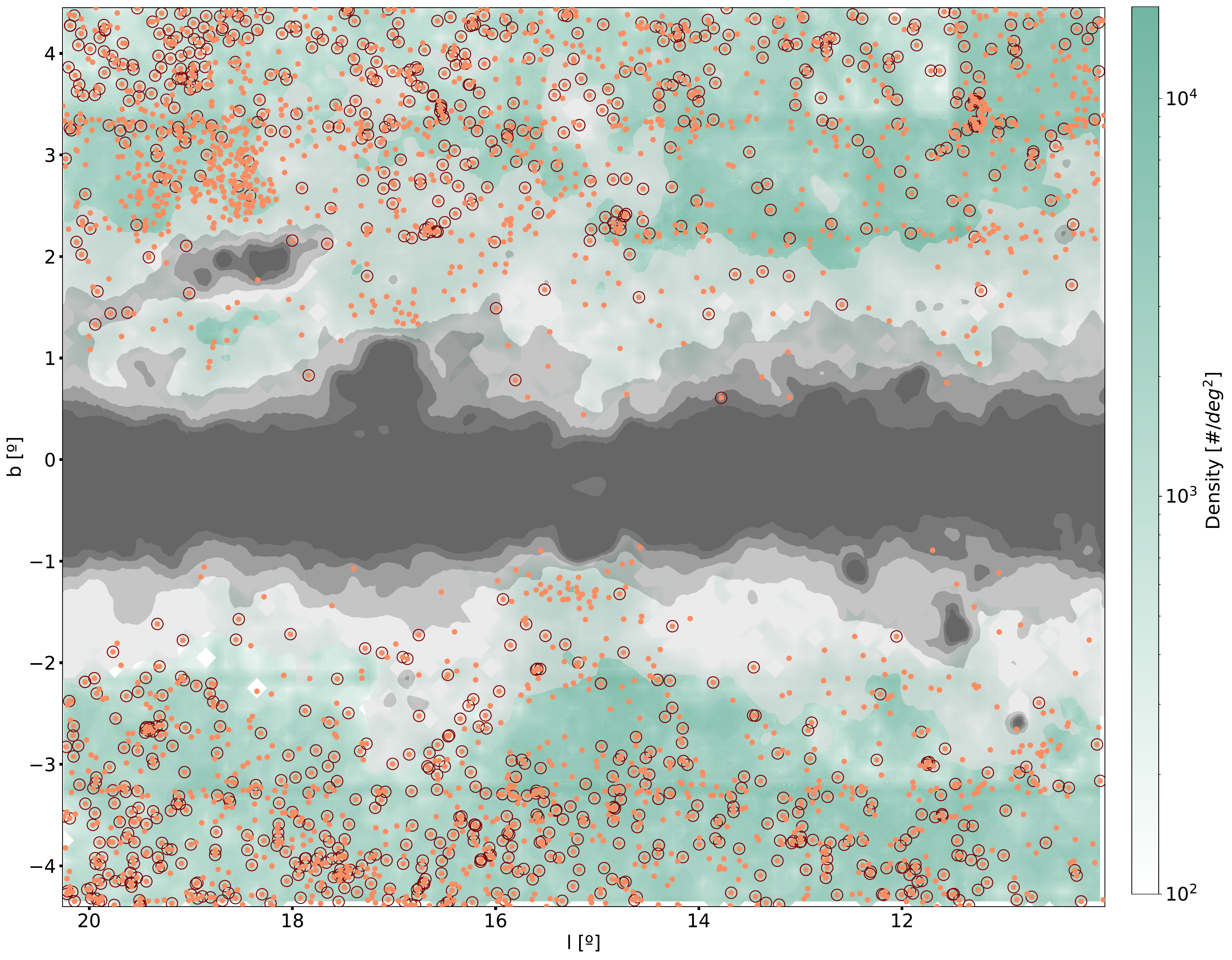}
\caption{Northern Galactic disc with Galactic longitudes between 20$^{\circ}$ and 10$^{\circ}$. Gxs are represented by orange points and non-Gxs as a density
green-scale map. Visual classifications of galaxies are shown as circles. The total optical A$_{V}$ interstellar extinctions from the maps of \citet{Schlafly2011} are superimposed in a grey gradient with levels of 10, 15, 20, and 25 mag.}
\label{fig:northerdisc}
\end{figure*}

Taking into account the trade-off between completeness and purity in the test set when the threshold is 0.6 for either of the two models, as described in section~\ref{sub:identification}, we obtained 2,818 Gxs selected from the probabilities of the possible extragalactic sources generated with the CNN and XGBoost  models. With visual inspection we determined that 1,003 are Gxs and 1,815 non-Gxs.  In Figure~\ref{fig:northerdisc}, we show the spatial distribution of the Gxs (orange points) and non-Gxs (density green-scale) classifications. The Gxs with visual inspection are represented by black circles. We superimposed the total optical Av interstellar extinction from the maps of \citet{Schlafly2011} such as Figure \ref{fig:CrossMatch}. 

\subsection{Analysis of the
two studied regions in the Galactic disc}

%IS:719 y PS:2692

\begin{figure}
    \centering
    \includegraphics[width=0.47\textwidth]{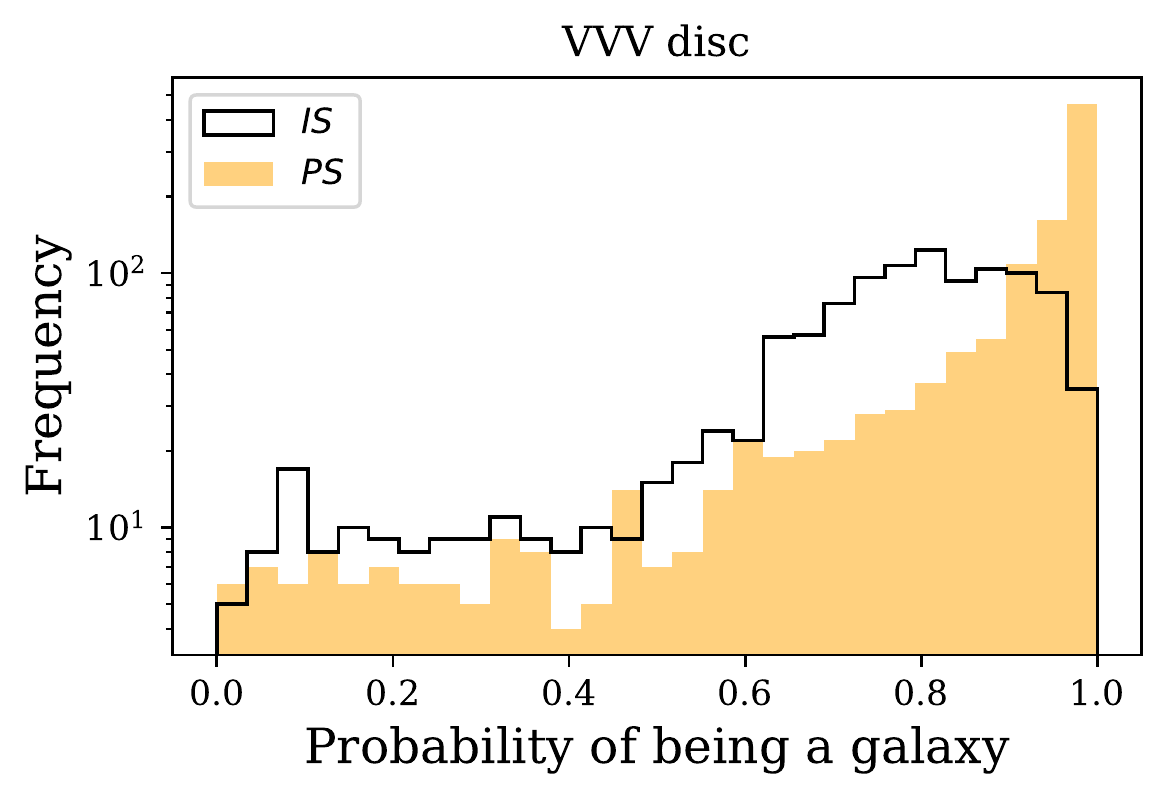}
    \includegraphics[width=0.47\textwidth]{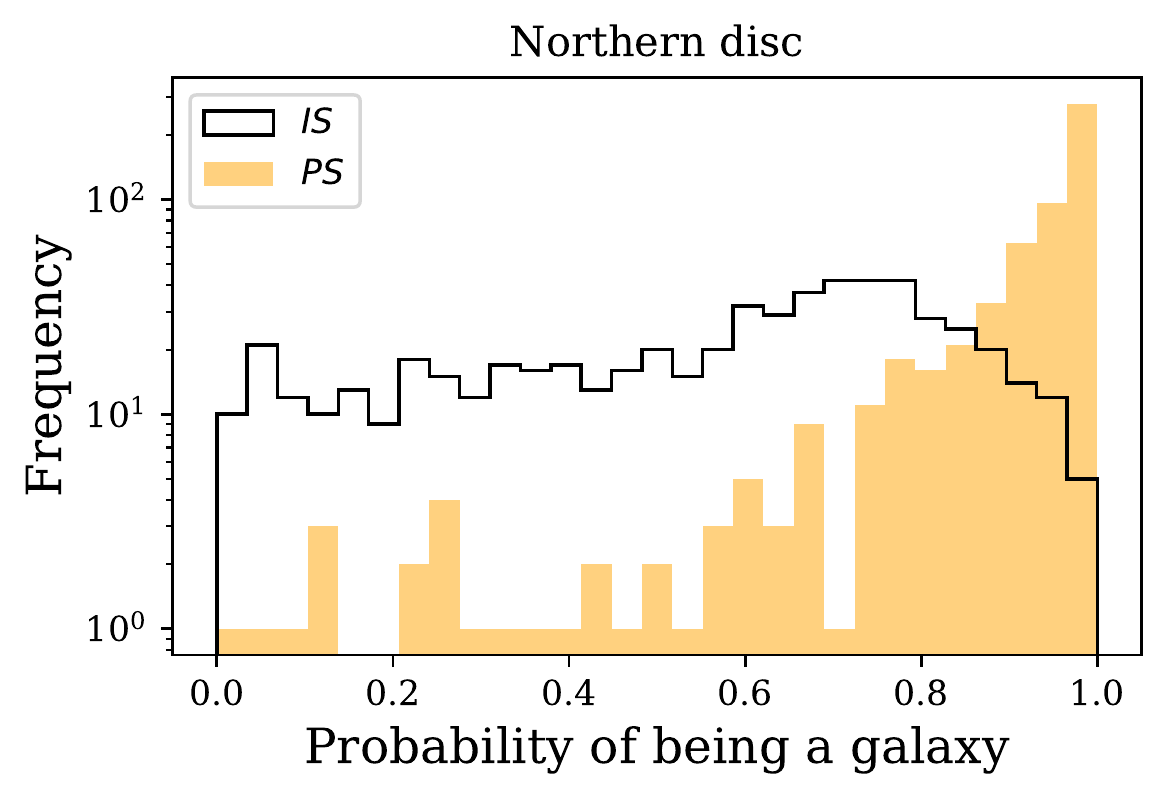}
    \caption{Probability distribution  of galaxies obtained of the set of 1,682 objects,  the probabilities of being a galaxy for the CNN model trained with IS in black and in orange the distribution of probabilities of being a galaxy for the XGBoost model trained with PS. The y-axis is shown in logarithmic scale. {\it Upper panel:} Galaxies belong to VVV. {\it Lower panel:} Galaxies belong to Northern disc.}
    \label{fig:VVV_disc_Northenr}
\end{figure}

We study the impact of implementing supervised models trained with data from different regions of the Galactic disc. The methodology chosen to create a catalogue of galaxies automatically subject to a decision threshold of 0.6 and the combination of using imaging and photometric/morphological information. We compare the precision metrics of Gxs obtained in VVV disc and Northern disc. We took two sets of 1,682 candidates to be galaxies that had probabilities greater than or equal to 0.6 for either model (CNN and XGBoost). One corresponds to a subset of the test set in the VVV disc with which we obtained a balance between the precision and recall of 0.65 and 0.69, respectively. The other set is a random subsample of objects in the Northern disc that satisfies the probability conditions. The probability distribution of galaxies for each model in the VVV disc and Northern disc regions are shown  in Figure~\ref{fig:VVV_disc_Northenr}.  Comparing the distributions in Figure~\ref{fig:IP_SP_VVV} and the distributions in the upper panel of Figure~\ref{fig:VVV_disc_Northenr}, we find that choosing a threshold greater than and equal to 0.6 for any of the models decreases the number of galaxies with probabilities close to zero compared to the distributions obtained when using the models individually and with a decision threshold of 0.5. Regarding the comparison between the distributions in the upper and lower panel of Figure~\ref{fig:VVV_disc_Northenr}, i.e. between objects under the same conditions but different regions, we find that the behaviour of the probabilities of galaxies in the Northern Disc are similar to those of galaxies in VVV greater than 0.6. From the visual inspections and model predictions we obtained the precision in these two sets when combining the two models and using them individually, the results are summarised in Table~\ref{tab:precision}. The results reflect that in the Northern disc region the use of photometric information affects the purity of the automatically generated catalogues with the XGBoost model trained with PS from 0.71 to 0.36. On the other hand, using only image data returns a similar purity to that obtained in VVV disc but decreases the number of Gx that can be obtained in this region compared to the number of galaxies obtained with the XGBoost model.

\begin{table}
    \caption{Result of the metric precision of the Gx classifications, for probabilities greater than or equal to 0.6 when both models or only one of them is used.}
    \centering
    \begin{tabular}{lcc}
    \hline
    Strategy & Precision  & Precision\\
             & VVV disc   & Northern disc\\
    
    \hline
    CNN $\cup$ XGBoost    & $\frac{1140}{1140+542}=0.68$ & $\frac{631}{631+1095} = 0.35$ \\
    CNN               & $\frac{950}{950+266}=0.78$   & $\frac{324}{324+100} = 0.76$\\
    XGBoost           & $\frac{1007}{1007+418}=0.71$ & $\frac{565}{565+1043} = 0.36$\\
    \hline
    \end{tabular}
    \label{tab:precision}
\end{table}

\begin{figure}
    \centering
    \includegraphics[width=0.5\textwidth]{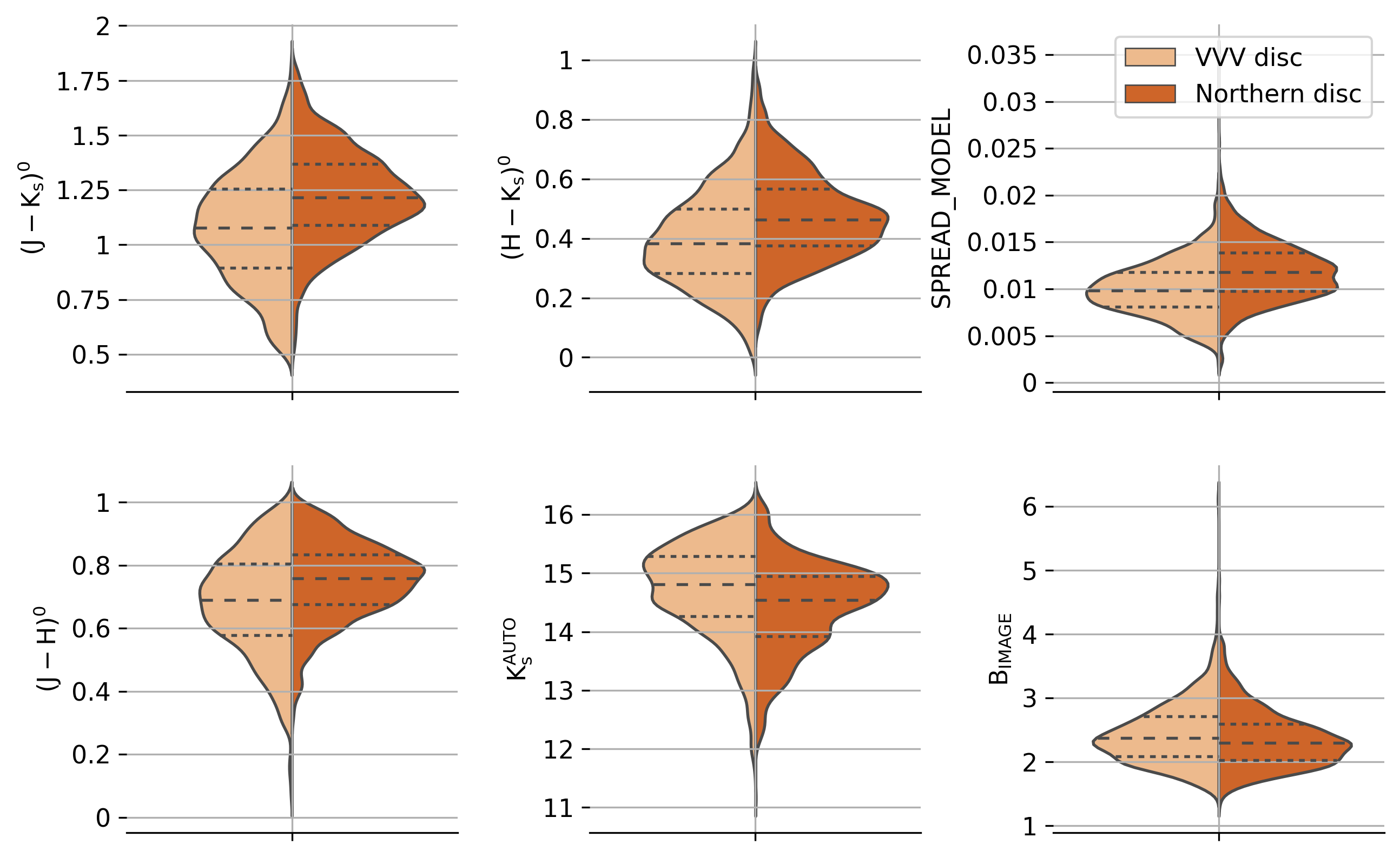}
    \caption{Kernel density estimation of PS sample features. The distributions correspond to galaxies in the VVV disc and the Northern disc. The dashed lines mark the sample medians and the dotted lines are the confidence intervals.}
    \label{fig:PS_gx}
\end{figure}

We compare the medians of the $J$, $H$ and $K_s$ bands and the photometric and morphological properties of 1,003 galaxies obtained by visual inspection in the Northern disc and randomly chose in  the VVV disc. In Table~\ref{tab:medians_images_79700} we show the median of the intensities of the images  and in Figure~\ref{fig:PS_gx} we show 6 of the most important features for the non-Gx and Gx classification according to the PCA method.

While the medians of the galaxy images are different as shown in Table~\ref{tab:medians_images_79700}, we find that the medians of the properties are statistically equal. Although the behaviours of the distributions between these galaxies are similar to those observed in VVV disc between non-Gx and Gx except for $B_{\textrm{IMAGE}}$, which is a property that has a fairly similar median and behaviour in the two regions and has to do with the shape of the objects. These comparisons may explain the effect of obtaining better accuracy with CNN and not with XGBoost on images and photometric and morphological data, respectively.

Considering the comparison between the non-Gx and Gx features of the candidates to be galaxies in VVV disc and the Northern disc (Table~\ref{tab:medians_images_79700}, Table~\ref{tab:properties}) and the comparisons in these same regions but only of visually classified galaxies such that the probabilities are greater than 0.6 for one of the CNN and XGBoost models (Table~\ref{tab:medians_images_79700}, Figure~\ref{fig:PS_gx}). We found that the differences in properties are mainly due to Gx candidates that were not taken into account in the VVV disc training.

\subsection{Catalogue}\label{Catalogue}

 We used the visual inspection of 2,818 Gxs and we performed a cross-match of 1~arcsec equatorial coordinates with sources on the SIMBAD Astronomical database\footnote{\href{https://simbad.unistra.fr/simbad/}{SIMBAD}}. As a result, out of these 2,818 Gxs, 1,815 sources were visually classified as non-Gxs and thus removed from the  catalogue, of which four are  Planetary Nebula (PN M 1-32, PMN J1821-2110, PMN J1757-1728 and PMN J1802-1909, \citep{2003yCat.2246....0C, 2011ApJS..194...25I, 1998AJ....115.1693C}), one dusty clump \citep{2013A&A...549A..45C} and one molecular maser \citep{2019ApJ...885..131R} found in SIMBAD. This procedure yielded a total of 1,003 galaxies visually confirmed with DSH J1827.0-2031 \citep{2012yCat.2311....0C} and ZOA J180953.827-123353.78 \citep{2014MNRAS.443...41W} previously classified as galaxies, the latter with spectroscopic information.

The VVV disc near-IR galaxy catalogue in the Northern part of the Galactic disc is available in electronic format \footnote{\href{https://catalogs.oac.uncor.edu/vvvx_nirgc/}{NorthernGalacticDisc\_catalogue.csv}}. The catalogue contains the identification in column (1), the J2000 equatorial coordinates in columns (2) and (3), the Galactic coordinates in columns (4) and (5), the A$_{Ks}$ interstellar extinction in column (6), total extinction-corrected $J^{0}$, $H^{0}$, and $K_{s}^{0}$ magnitudes in columns (7) to (9), the extinction-corrected $J_{2}^{0}$, $H_{2}^{0}$, and $K_{s}$ $_{2}^{0}$ aperture magnitudes within a fixed aperture of 2~arcsec diameter in columns (10) to (12), the morphological parameters: $R_{1/2}$, $C$, ellipticity and $n$ in columns (13) to (16), the probabilities of the IS-CNN and PS-XGBoost models in columns (17) and (18), respectively. All magnitudes were transformed to the 2MASS photometric system. In Table~\ref{tab:catalogue} we show the properties of the first five galaxies of the Northern part of the Galactic disc catalogue. The 1,003 extragalactic galaxies included were obtained with probabilities greater than 0.6 resulting from the CNN or XGBoost models. 

\begin{table*}
\caption{The VVV NIRGC: Northern part of the Galactic disc. The full table is available online.}
\begin{tabular}{lccccccccc}
\hline
ID & $\alpha$ (J2000) &  $\delta$ (J2000) & l & b & A$_{Ks}$ & $J^{\textrm{AUTO}}$  &  $H^{\textrm{AUTO}}$ & $K_s^{\textrm{AUTO}}$ &  $J_2$ \\
 & [$^{\circ}$] &  [$^{\circ}$] & [d$^{\circ}$ & [$^{\circ}$] &  [mag]  &  [mag] &  [mag]  &  [mag]  &   [mag]\\
\hline
VVVX-J175153.74-180752.9 &   267.974 &   -18.131 &  10.015 &  4.323 &  0.301 &   15.619 &   15.125 & 14.803 &   16.782\\

VVVX-J175156.54-180638.2 &   267.985 &   -18.111 &  10.039 &  4.323 &  0.297 &   15.293 &   14.599 & 14.382 &   16.478\\

VVVX-J175204.21-180553.7 &   268.017 &   -18.098 &  10.065 &  4.304 &  0.296 &   14.864 &   13.852 & 13.813 &   15.854\\

VVVX-J175211.72-175146.1 &   268.049 &   -17.863 &  10.283 &  4.397 &  0.274 &   15.228 &   14.401 & 14.154 &   16.093 \\

VVVX-J175224.46-180414.1 &   268.102 &   -18.070 &  10.130 &  4.248 &  0.287 &   15.157 &   14.202 & 13.931 &   15.996 \\
\hline
                    &              &              &
&           &           &             &             &
          &             \\
\hline
 &  $H_2$ &  $K_s{}_2$ & $R_{1/2}$ & C &  $e$  & $n$ &  prob\_IS &  prob\_PS\\
 &   [mag] &   [mag]  &  [arcsec]  &  &   &  &   &  \\
\hline
   &   16.067 &    15.716 &  1.239 &  3.147 &      0.457 &  4.095 &    0.382 &    0.613\\

   &   15.679 &    15.305 &  1.150 &  2.459 &      0.290 &  2.101 &    0.495 &    0.972\\

   &   15.156 &    14.714 &  1.163 &  3.125 &      0.394 &  3.689 &    0.703 &    0.996\\

 &   15.290 &    15.025 &  1.076 &  2.161 &      0.280 &  4.091 &    0.277 &    0.810\\

 &  15.317 &    14.880 &  1.165 &  2.303 &      0.354 &  4.773 &    0.360 &    0.959\\
\hline
\end{tabular}
\label{tab:catalogue}
\end{table*}

\section{Main conclusions}\label{sec:conclusions}

We performed an intensive research for the automatic generation of galaxy catalogues at low-latitude regions of the Milky Way with high densities of stellar objects, gas, and dust. To carry out this task, we used observations from the VISTA Variables in the Vía Láctea \citet[VVV]{Minniti2010} and the extended survey \citet[VVVX]{Minniti2018}. Following the methodology implemented in \cite{Baravalle2018}, cleaning point sources from cross-match with Gaia and removing sources at the edge of images, we obtained possible extragalactic sources which have image information in the $J$, $H$ and $K_s$ passbands (Image Sample, IS), in addition to photometric and morphological information (Photometric Sample, PS). We performed all the analysis and training of the machine learning algorithms using the  information coming from the VVV
NIRGC catalogue \citep{Baravalle2021}. The VVVx data were used as a case study for the automatic generation of galaxy catalogs with the procedure we developed in this work.

For IS, we found that for both unsupervised and supervised methods, scaling the images by passband using the intensity ranges of the $J$, $H$ and $K_s$ passbands provides the most relevant information for galaxy identification. Nevertheless, we found that for the unsupervised methods the images with higher entropy are those with spatial size 11 $\times$ 11 pixels, while for the supervised methods the best results were obtained with an image spatial size of 44 $\times$ 44 pixels. This is due to the fact that the latter contains the source in the central part and also background information of the sky, such as stars. In addition, we found that adding information from the edges of the detected sources in each passband through the Gauss-Laplace filter generates an improvement in the performance of the Convolutional Neural Network (CNN), reaching a performance of $F_1=0.68$ measured on the Gxs class in the  test set. 

As for PS, the best performing model corresponds to Extreme Gradient Boosting (XGBoost) trained with 22 morphological and photometric features. The model was trained without class balancing, achieving an $F1 = 0.65$ on the test set. A similar metric value was achieved employing Neural Networks, however, we prefer the former model as it obtained a better recall metric.
Additionally, we explored the performance of the models using 7 selected features with Mutual Information and performing a Principal Component Analysis. In both cases, the performances obtained were slightly lower. Nevertheless, this feature selection methodology is useful when one wants to apply these models in another study with few computational resources. 

Comparing the classifications made in VVV disc by the CNN and XGBoost models, we observed a similar behaviour in the Recall-Precision curves. In general, the results obtained with CNN in IS were better. In order to enhance the classification of the two approaches (IS-CNN and PS-XGBoost), we combined both galaxy classification methods. We selected those candidates to be galaxies that have probabilities greater than 0.6 in either of the two models. With this choice we obtained a total F1 score of 0.67 on the test set, and a balance between precision and recall of 0.65 and 0.69, respectively.  This gives a similar percentage of galaxies as that obtained with the visual inspection in \cite{Baravalle2021}.

We have designed an automatic methodology that includes the improved pipeline which selects possible extragalactic sources, two supervised automatic learning models for the classification of Gxs and non-Gxs and the criteria for the generation of automatic galaxy catalogues. We applied these models to the 172,396 possible extragalactic sources obtained in the Northern part of the Galactic disc using the VVVX survey. Of these candidates, 2,818 Gxs were automatically classified by the combination of IS-CNN and PS-XGBoost models.

In order to analyse the classifications of the models, we take two subsets in the VVV disc and the Northern disc of 1,682 candidates under the criteria of likelihoods greater than 0.6 for any model with visual inspection. We find that the balance between precision and recall in the VVV disc is not recovered in the Northern disc when the models are combined. When analysing the probabilities of the models individually, we find that the CNN model trained with IS has similar accuracy to that obtained in VVV disc and is more robust than the XGBoost model trained with PS. However, the automatic catalogue generated only with CNN-IS finds fewer galaxies than those obtained with XGBoost.

% global differences, differences between galaxies
The features / pixels generated with edge-on filters as part of the input to a CNN are important for the classification between GXs and non-GXs. Since the intensities in the $J$, $H$ and $K_s$ bands are very different between the VVV disc and Northern disc regions, both in the candidate set of 79,700 objects and in the comparison between galaxies. Therefor, the pixel intensities used are not the most important feature in this region for classification.

On the other hand, the statistical differences in the median photometric and morphological properties of all VVV disc and Northern disc candidates are larger than those obtained when studying the galaxies separately, so the PS-XGBoost model did not have candidate examples in the learning, which affects its result.

Finally,  we visually inspected the 2,818 Gxs automatically classified generating the catalogue of 1,003 galaxies in the Northern disc, which we make publicly available in this work. In the subsection~\ref{Catalogue} it is shown that  these galaxies have photometric and morphological parameters that are consistent with early-type galaxies, similar to those catalogued in the VVV NIRGC. This effect could be due to the fact that the interstellar extinction in these regions is slightly lower than in the previously studied regions, where we could only observe the galaxy bulges. Our catalogue contains only two previously reported galaxies, although only one of them has an available radial velocity.

In the future, we intend to apply this methodology to the other areas of the VVVX survey, which covers a wide region of the ZoA, poorly explored in extragalactic terms. Only with spectroscopic data we would be able to confirm the extragalactic nature of these objects.

%%%%%%%%%%%%%%%%%%%%%%%%%%%%%%%%%%%%%%%%%%%%%%%%%%%%%%%%%%%%%%%%%%%%%%%%%%%%%%%%%%%%%%%%%%%%%%%%%%%%%%%%%%%%%%%%%%%%%%%%%%%%%%%%%%%%%%%%%%%%%%%%%%%%%%%%%%%%%5
\section*{Acknowledgements}

We would like to thank the anonymous referee for the useful comments and suggestions which has helped to improve this paper.

This work was partially supported by Consejo de Investigaciones Cient\'ificas y T\'ecnicas (CONICET) and Secretar\'ia de Ciencia y T\'ecnica de la Universidad Nacional de C\'ordoba (SeCyT). 
The authors gratefully acknowledge data from the ESO Public Survey program IDs 179.B-2002 and 198.B-2004 taken with the VISTA telescope, and products from the Cambridge Astronomical Survey Unit (CASU).
J. L. N. C. is grateful for the financial support received from the Southern Office of Aerospace Research and Development (SOARD), a branch of the Air Force Office of Scientific Research's International Office (AFOSR/IO) through grants FA9550-22-1- 0097.

%%%%%%%%%%%%%%%%%%%%%%%%%%%%%%%%%%%%%%%%%%%%%%%%%%
\section*{Data Availability}
  
The VVV NIRGC used in this article is available at
\url{https://catalogs.oac.uncor.edu/vvv\_nirgc/}. 
This list of probabilities of possible extragalactic sources can be used to generate catalogues according to the case study to be performed and is available electronically,  as is
the VVV near-IR galaxy catalogue: Northern part of the Galactic disc at \url{https://catalogs.oac.uncor.edu/vvvx_nirgc/}.

%%%%%%%%%%%%%%%%%%%% REFERENCES %%%%%%%%%%%%%%%%%%

% The best way to enter references is to use BibTeX:

\bibliographystyle{mnras}
\bibliography{Bibliography} % if your bibtex file is called example.bib

%%%%%%%%%%%%%%%%%%%%%%%%%%%%%%%%%%%%%%%%%%%%%%%%%%

%%%%%%%%%%%%%%%%% APPENDICES %%%%%%%%%%%%%%%%%%%%%

\appendix

\section{Models}\label{app:TheBetsModels}

The supervised machine learning models CNN and XGBoost were trained with IS and PS, respectively. These models achieve the best performance with a customised configuration of some parameters and hyper-parameters. 

In order to make the methodology implemented in this work reproducible, details of each of the models are given below. 

\begin{figure}
    \centering
    \includegraphics[width=0.28\textwidth]{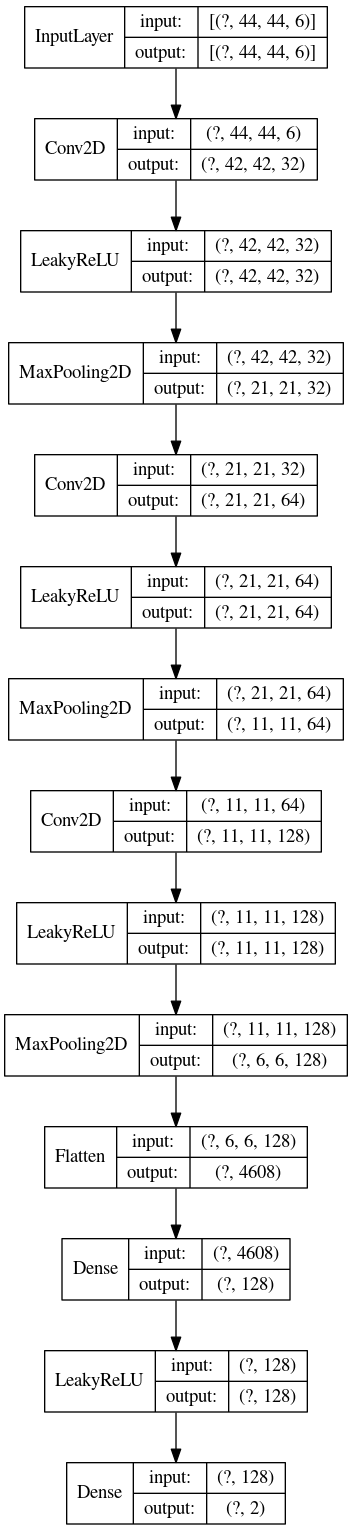}
    \caption{Sequence and shape of CNN layers. The question mark indicates the number of images entering the network.}
    \label{fig:CNN_keras}
\end{figure}

The CNN model was performed with the \textsc{keras} library\footnote{\href{https://keras.io/}{Keras library}} using \textsc{TensorFlow} as backend. The neural network is composed of three Conv2D layers with kernels of (3, 3), three MaxPooling2D with pool\_size iqual to (2, 2), four activation functions Leaky Relu and one activation function softmax. In total, the number of parameters used in the training was of 684,322 with batch\_size $= 35$ and 40 epochs. In Figure~\ref{fig:CNN_keras} we show the design of the network

The gradient boosting algorithm used in this work is the implementation given in the open source package \textsc{XGBoost}\footnote{\href{https://xgboost.readthedocs.io/en/stable/}{XGBoost library}}. The tuning of the model hyper-parameters was carried out by optimising the cross-validation F1 score using an evolutionary algorithm from the \textsc{DEAP} package\footnote{\href{https://sklearn-genetic-opt.readthedocs.io/en/stable/andindex.html}{Sklearn genetic opt}}. The set of parameters giving the best performance in classifying Gx and non-Gx objects are shown in Table~\ref{tab:XGBoost_table}.

\begin{table}
    \caption{Fitted parameters of the XGBoost model.}
    \centering
    \begin{tabular}{lc}
    \hline
    Parameters &  Values\\
    \hline
    objective          & binary:logistic\\
    colsample\_bytree   & 0.84\\
    gamma               & 1e-3\\
    learning\_rate      & 0.46\\
    max\_depth          & 15\\
    min\_child\_weight  & 4\\
    n\_estimators       & 100\\
    reg\_alpha          & 1e-05\\
    reg\_lambda         & 50.0\\
    scale\_pos\_weight  & 13.48\\
    subsample           & 0.93\\
    eval\_metric        & aucpr\\ 
    \hline
    \end{tabular}
    \label{tab:XGBoost_table}
\end{table}

% Don't change these lines
\bsp	% typesetting comment
\label{lastpage}
\end{document}